\documentclass{./styles/svproc}
\usepackage{url}

\usepackage[english]{babel}
\usepackage{graphicx}
\usepackage{subcaption}
\usepackage{appendix}
\usepackage{hyperref}
\usepackage{booktabs}
\usepackage{subfiles} 
\renewcommand{\arraystretch}{1.5}

\usepackage{array}
\usepackage{hhline}
\usepackage{booktabs}
\usepackage{rotating}
\usepackage{amssymb}
\usepackage{changepage}
\usepackage{makecell}
\usepackage{pifont}

\usepackage{tablefootnote}
\usepackage{multicol, multirow}
\usepackage{pdflscape} 

\usepackage{booktabs}
\usepackage{arydshln}
\usepackage{tabulary}
\usepackage{multirow}
\usepackage{multicol}
\usepackage{todonotes}

\newcommand{\blue}[1]{{\color{black}#1}} 

\usepackage{lineno}

\usepackage{mathtools}
\usepackage[sorting=none]{biblatex} 

\begin{document}
\mainmatter              
\title{Structure and dynamics of growing networks of Reddit threads}
\titlerunning{Structure and dynamics of growing networks of Reddit threads}  

\author{Diletta Goglia\inst{1} \and Davide Vega\inst{1}}
\authorrunning{Diletta Goglia \and Davide Vega} 
%
\tocauthor{Diletta Goglia, Davide Vega}
\institute{\blue{Infolab,} Department of Information Technology, Uppsala University, Uppsala, Sweden\\
\email{\{diletta.goglia,davide.vega\}@it.uu.se}
}

\maketitle 

\begin{abstract}
Millions of people use online social networks to reinforce their sense of belonging, for example by giving and asking for feedback as a form of social validation and self-recognition.
It is common to observe disagreement among people beliefs and points of view when expressing this feedback. Modeling and analyzing such interactions is crucial to understand social phenomena that happen when people face different opinions while expressing and discussing their values.
In this work, we study a Reddit community in which people participate to judge or be judged with respect to some behavior, as it represents a valuable source to study how users express judgments online. 
We model threads of this community as complex networks of user interactions growing in time, and we analyze the evolution of their structural properties. We show that the evolution of Reddit networks differ from other real social networks, despite falling in the same category. This happens because their global clustering coefficient is extremely small and the average shortest path length increases over time. Such properties reveal how users discuss in threads, i.e. with mostly one other user and often by a single message. 
We strengthen such result by analyzing the role that disagreement and reciprocity play in such conversations.
We also show that Reddit thread's evolution over time is governed by two subgraphs growing at different speeds. We discover that, in the studied community, the difference of such speed is higher than in other communities because of the user guidelines enforcing specific user interactions.
Finally, we interpret the obtained results on user behavior drawing back to Social Judgment Theory.

\end{abstract}

\section{Introduction}
\label{sec:intro}

Conversations \blue{on} online social media have become a crucial aspect of modern communication~\cite{smith_chapter_2018}, shaping how individuals interact with each other, share information, and form connections~\cite{Grabowicz2011SocialFO}. Social network platforms enable conversations to reach a wide audience, \blue{while} allow\blue{ing} for real-time sharing of information, opinions, and reactions. These online conversations are driven by three key factors: (i) the purpose of the users and how they want to communicate, (ii) the functionality provided by the platform and its limitations, and (iii) the user guidelines and recommendations governing the community~\cite{book_custodians, russo_spillover_2023}. The later two factors, functionality and guidelines, are key to encourage \blue{certain} types of interactions and conversational structures \blue{over} others. For example, while X (formerly known as Twitter) used to enforce users to write messages with less than 140 characters, Facebook and other thread-based forums do not have such \blue{restrictions}, leading to more informative messages and shorter interactions~\cite{alis_quantifying_2015}. \blue{Additionally}, some platforms have guidelines and recommendations that, \blue{while} not \blue{technologically} enforced\blue{, aim} to guide user behavior. These guidelines serve as suggestions encouraging users to conform to an expected conduct. \blue{For instance,} Instagram\blue{'s} Community Guidelines\footnote{\url{https://help.instagram.com/477434105621119}} promot\blue{e} authentic interactions to avoid spam, \blue{and} Facebook groups \blue{often have customized rules}.
Whether these guidelines \blue{effectively} contribute to shap\blue{e} user behavior \blue{remains} to be explored.
It is well known, \blue{however}, that many of these rules, regulations and technical limitations change over time \blue{based on users interactions} with the platform. 
A recent example \blue{of} how users have forced platforms to modify their regulations is X. Before 2018, when users needed to write longer tweets (i.e. with more than 140 characters) \blue{would} split the long text into multiple interconnected tweets\blue{,} including formats like ``Tweet 1/12''. In \blue{December 2017}, X adapted \blue{to this behaviour by} introduc\blue{ing} a \blue{new} feature called threads, allow\blue{ing} users to \blue{concatenate multiple tweets together in a sequence. This made easier to create and follow longer conversations or narratives within a single thread.}

The goal of this work is to uncover the effect of platform guidelines on online conversations and evaluate \blue{the extent} to wh\blue{ich} such guidelines influence participation. In particular, we focus on Reddit, a social media platform where people participate in ``self-governing and self-organized'' communities \blue{known as} \textit{subreddits}~\cite{jamnik_use_2019, medvedev_anatomy_2019}\blue{. E}ach \blue{subreddit has its} own rules and guidelines\blue{, specifying} what is allowed inside the subreddit and recommend\blue{ing} how \blue{users should} behave, often based on a specific topic (e.g., \texttt{r/science}, \texttt{r/gaming}). Beyond \blue{being} simple forum, Reddit has been widely populated with subreddits with their own guidelines and internal features, making it a valuable resource for conducting social research on opinion formation~\cite{shatz_2017, hintz_2022}. \blue{In recent years, a plethora of studies of interesting subreddits has emerged, particularly focusing on studying their evolution and dynamics to understand how such communities develop and grow over time~\cite{krohn2019modelling, weninger_discussion_threads, horawalavithanaOnlineDiscussionThreads2022}.
Many studies has also analyzed discussions in Reddit communities to understand how interactions among participants influence behavior. For example, Petruzzellis et al. exploited the \texttt{r/ChangeMyView} subreddit to analyze changes in online information consumption behavior arising after opinion changes~\cite{opinion_change_reddit_morales}. In~\cite{CAUTERUCCIO2023103516}, Cauteruccio et al. investigated the emotional experiences in eSports spectatorship using the \texttt{r/leagueoflegends} subreddit: they show that spectators supporting the same team tend to engage in cohesive discussions, while interactions among those supporting different teams are less salient.
Additionally, a significant body of research has focused on the language used in Reddit discussions, examining linguistic patterns, sentiment, and rhetorical strategies to gain deeper insights into the nature and impact of online communication within these communities.
For instance, Helm et.al.~\cite{incel_reddit} studied the \texttt{r/Incel} community to identify subcultural discourse and understand how it affirms deviant behaviors, while Bouzoubaa et al.~\cite{drugs_reddit} analyzed drug-related subreddits to understand their role in the online discourse surrounding substance use.}

\blue{O}n Reddit, users can write and publish posts (\blue{known as} \textit{submissions}) or comments\blue{. E}ach post, together with its comments, constitutes a \textit{thread}, i.e., a conversation where comments are organized hierarchically in a tree-like format. The post is the root of the thread, each comment is a node in the discussion tree, and replies to post\blue{s} or comment\blue{s} create branches in the tree. Reddit communities are moderated by designated users (the \textit{moderators}) who establish the community rules and ensure that everyone follows them when participating. Moderators \blue{maintain} order \blue{by performing} actions\blue{ such as} deleting posts or comments \blue{and} banning users. In each community, the rules \blue{are} often \blue{displayed} on the side of the webpage.

In this work we focus our attention on the \texttt{/r/AmItheAsshole} (AITA) subreddit\footnote{\url{https://www.reddit.com/r/AmItheAsshole}\\ According to~\cite{2023_reddit_recap}, AITA is the most viewed Reddit community since 2020.}, an online community \blue{where} people post stories about personal experiences having ambiguous moral valence\blue{,} asking othe\blue{s} if they have been ``assholes" \blue{(}or not\blue{)} in the narrated story, i.e. if they are to blame for the conflict \blue{described}. Users creating such posts should provide detailed descriptions of their stories in the text, including relevant background information about the people involved. \blue{Other users then perform} \blue{t}he explicit judgment by \textit{voting}, \blue{which involves} writing a comment including a specific acronym corresponding \blue{their} judgment. The available acronyms provided by the community are listed in Table \ref{tab:flair_acronyms}.

The subreddit guidelines suggest that, \blue{along} with the acronym, users should include in the comment a brief motivation for the vote \blue{to explain} their choices to other readers. The AITA community \blue{uses} Reddit's integrated voting system to allow participants to rate the judgments they agree with by upvoting them. Expressing disagreement is not allowed in this context, since downvotes are used to report off-topic or spam discussions and harassing comments. The community \blue{has} established a\blue{n} 18-hours waiting period before assigning the final verdict. Users must vote within this timeframe. \blue{As} users upvote different comments, a consensus emerges over time, with one judgment gaining \blue{the majority of} agreements as the collective decision. After the time window passes, this judgment is then accepted as the official verdict and is made public by assigning a flair to the post, i.e. a tag with the respective judgment acronym. More details about the voting process are provided in Section \ref{subsec:operationalization}.

\renewcommand{\arraystretch}{1.3}
\setlength{\tabcolsep}{6pt}
\begin{table}[h!]
  \centering
  \small
  \begin{tabular}{ p{2.8cm}  p{4cm} p{4.6cm} }
  \toprule
  
  \textbf {Acronym} & \textbf{\makecell{Corresponding\\judgment}} & \textbf{Meaning} \\ \cdashline{1-3}
  
  \toprule

YTA or YWBTA & ``You're the Asshole" or ``You Would be the Asshole'' & The author of the post is the one behaving immorally in the story. \\ \cdashline{1-3}
NTA or YWNBTA & ``Not the Asshole" or ``You Would Not be the Asshole'' & The author of the post did not behave immorally in the narrated story. \\ \cdashline{1-3}
ESH & ``Everyone Sucks Here'' & All the characters of the narrated story behaved immorally. \\ \cdashline{1-3}
NAH & ``No A-holes Here'' & None of the characters behaved immorally. \\ \cdashline{1-3}
INFO & ``Not Enough Info'' & There are missing details needed to express a judgment. \\
  \bottomrule
  \end{tabular}
  \\[\baselineskip]
  \caption{Acronyms provided by the AITA community.}
  \label{tab:flair_acronyms}
\end{table}

%

In the AITA community, the explicit request for a judgment is, therefore, a requirement of the subreddit, allow\blue{ing} researchers to study how humans express moral judgments through socio-linguistic features. Indeed, the comments contained in AITA threads offer the ground truth of what people voted for and often why. This motivates why the AITA community has received much attention the last two years. Botzer et al.~\cite{botzer_analysis_2023} exploited the AITA subreddit to study the presence and impact of moral valence, as well as whether gender and age play a role in users' judgments. De Candia et al.~\cite{de_candia_social_2022} \blue{examined} which demographic factors and topics are associated with judgment\blue{s}, while Giorgi et al.~\cite{giorgi_author_2023} analyzed the possibility of identifying, \blue{through} linguistic and narrative features, whether the author of the post is also the character \blue{in} the story or is narrating a story from a third\blue{-}person perspective.

In order to analyze the dynamics of interactions in the AITA subreddit, we collected more than 6,000 threads that received \blue{significant} attention in 2023 (see details in Section \ref{subsec:data}). \blue{For each thread, we compute the individuals' amount of judgment and the group level of disagreement (Section \ref{subsec:operationalization})}. \blue{We t}hen model each thread as a complex multi-graph network of user interactions evolving \blue{over} time \blue{(Section \ref{subsec:temporal_network})}. We study the growth of such networks reconstructing each conversation \blue{over} time and comparing the evolution of structural properties with respect to the exist\blue{ing} literature on growing real social networks\footnote{In this work we will use the term ``real social networks'' to refer to networks modeling social interactions which comes from real-world data, i.e. non-random and not synthetically generated}, including other subreddits \blue{(Section \ref{sec:results})}. In particular, we focus on the clustering coefficient and average shortest path length as structural properties growing \blue{over} time, since \blue{these} highlight the peculiar evolution of AITA networks and help explain the reason{s} behind \blue{the} user behavior. \blue{Furthermore, we compute the reciprocity and the disagreement of such networks to understand if they play a role in the AITA discussions.}

In short, the \textbf{contributions} of this work are the following:

\begin{itemize}

    \item Our temporal analysis of the communication exchange shows that Reddit user interaction networks consist of two subgraphs\blue{, a \textit{star} and a \textit{periphery},} that exhibit different speeds of growth \blue{(Section \ref{sec:results})}. \blue{The star structure is mainly formed by users not engaging in conversations and rather answering to the root message of each subreddit thread (i.e., post), while the periphery is mostly composed of users engaging in long conversations.}\\
    
    \item \blue{We find that the speed at which participants contribute in these subgraphs is highly influenced by the intention of the participants. In the periphery subgraph, the participants who vote in addition to writing comments respond almost twice as slowly as users just commenting (Section \ref{subsec:results_substructures})}. \blue{At a macro level, w}e explain how people engage in conversations with other users in subreddits \blue{through} the insights revealed by the evolution of structural properties. Specifically, the increasing average shortest path length as well as the decreasing (and very small) clustering coefficient \blue{reflect} the behavior of people discussing mostly with only one other user, often \blue{through} a single message \blue{(Section \ref{subsec:results_aita})}.\\

    \item \blue{Our analysis shows that these} interaction networks evolve differently \blue{compared} to other social networks, despite falling into the same category of ``real social networks'' \blue{(Section \ref{subsec:results_reddit})}. \blue{More specifically, w}e compare the AITA subreddit with other subreddits \blue{by examining} the growth of the \blue{two subgraphs within the} dynamic networks. We \blue{demonstrate} that the speed \blue{of the \textit{star} subgraph is between 2 and 3 times larger than \blue{that of the} \textit{periphery} subgraph} in AITA\blue{, which} is significantly larger \blue{than in other Reddit communities}. \blue{W}e interpret \blue{this} as a consequence of community rules shaping user behavior.\\

    \item Our analysis shows that disagreement in the judgment process is \blue{associated} with more interactions in the thread but \blue{may} prevent some users express\blue{ing} a judgment. Specifically, we prove that when the disagreement is higher (i.e.\blue{,} when the judgment is not obvious), people prefer to discuss rather than judge: they engage with others \blue{through} more comments\blue{,} and if they express a vote\blue{,} they \blue{struggle to} clearly pick a side (Section \ref{subsec:reciprocity}).\\
    
\end{itemize}

Finally, we analyze the underlying social dynamics among users \blue{by} drawing on social psychology theories and interpreting their effect on the graph structure evolution. Specifically, we interpret our results \blue{through the lens of} Social Judgment Theory~\cite{brehmer_chapter_1988}.

\section{Theoretical framework}
\label{sec:theoretical_framework}
One of the goals of this work is to shed light on how people discuss in online communities where they are asked to explicitly express their opinion. Specifically, we aim \blue{to} measure to what extent users' disagreement affects the evolution of the online conversation\blue{s}. We do this by modeling all the threads in the subreddit as a set of growing networks of user interactions (see Section~\ref{sec:method}). In this section, we lay the groundwork for understanding social judgment dynamics \blue{(Section \ref{subsec:judgment})}, and we provide the state of the art of growing social networks \blue{(Section \ref{sota:growing_net})}.  

\subsection{Social Judgment}
\label{subsec:judgment}
In social psychology, judgment is defined as the cognitive process of forming opinions, evaluations, or assessments about oneself, others, or situations. It generally consist in the product of non-conscious systems that operate quickly \blue{based on} some evidence \cite{gilbert_2002}. For example, when engaging in a conversation with someone, body language, tone of voice, and facial expressions are cues that serve as evidence to formulate judgments about the person. Social psychologists have studied various aspects of judgment, including how people make decisions, evaluate others, and interpret social information. Specifically, Social Judgment Theory (SJT)~\cite{brehmer_chapter_1988} is a theoretical framework within social psychology that seeks to understand how individuals form and evaluate judgments about themselves and others. SJT also investigates the reasons why, in particular social contexts, people are more inclined to express judgments~\cite{morrison_distinguishing_2008, noelle-neumann_spiral_1993, morrison_explaining_2011, hornsey_2003, matthes_morrison_2010}. For example,~\cite{Adamic_2021, spears_social_2021} found that users are more prone to express negative judgments in anonymous settings where either the giver or the receiver of the opinion \blue{is} unknown. Despite the \blue{extensive} scientific literature, we have little understanding about the role that disagreement plays in such settings.

Research on user interactions in online platforms has primarily \blue{focused} on conflict, controversies, and affective polarization~\cite{addawood-etal-2017-telling, 10.1145/3140565, 7403517, Mejova_2014, Conover_Ratkiewicz_Francisco_Goncalves_Menczer_Flammini_2021}, analyzing these social behaviors mostly through sentiment and topic analysis.
In particular, Kumar et al.~\cite{10.1145/3178876.3186141} used Reddit data to study conflictual interactions of users across different communities. They found that less than 1\% of communities start the majority of conflicts and that such conflicts are initiated by highly active community members and carried out by significantly less active members. In our work, instead, we are interested in studying the role of disagreement among users.
Despite the plethora of studies about the role of polarization and conflict in online conversations, \blue{the question of} if and how disagreement affects \blue{people's} moral judgments \blue{remains} unexplored.

\subsection{Growing social networks }
\label{sota:growing_net}

Conversational data, such \blue{as} the actions and interactions of users in online platforms, can be modeled as dynamic social networks~\cite{newman_structure_2003, scott_sna, wasserman_social_1994}. For example, a follower-followee relationship\blue{s}, Facebook friendship links, e-mail or message exchanges, and retweet patterns. When these networks are not synthetic but taken from real user interaction data, they are commonly referred to as ``real social networks"~\cite{newman_structure_2003, leskovec_graphs_2005} to emphasize that the original data \blue{originates} from actual networks rather than mechanistic models. These types of networks include a \blue{wide} variety of online connections such as friendship or following relations in social media (e.g., X), interactions such as sharing message\blue{s}, replying to emails, or real-life interactions (e.g., academic co-authorship). The properties of this category of networks have been extensively studied \blue{from} both a static and a dynamic \blue{perspective}. The structural evolution of growing social networks has been intensively studied by Newman, who analyzed structural properties of some models of growth~\cite{newman_structure_2003}, proved that preferential attachment is the origin of power-law degree distributions in collaboration networks~\cite{newman_pref_2001}, and developed a new growing model that reproduces features of real-world friendship networks~\cite{newman_growing_2001}. However, most \blue{research has} focus on study\blue{ing} structural properties of networks after a sufficiently \blue{long} period, rather than on how such properties evolve during networks' growth (See Table~\ref{tab:sota_growing_net}). An exception is the work of Leskovec~\cite{leskovec_graphs_2005} who, \blue{through} empirical observation of four real graphs (three of which were social) growing \blue{over} time, demonstrate\blue{d} that such networks become denser over time and that their diameter shrinks.

In summary, \textit{real social networks} \blue{represent} a subclass of social networks that includes a \blue{wide} variety of graphs with diverse underlying dynamics. As a result, discoveries in the literature about \blue{the growth of} real social network \blue{structures and properties} over time may not be \blue{universally} applicable to all graphs \blue{within} this class. For example, it is reasonable to think that a graph of retweets could grow \blue{differently over} time \blue{compared to} graph of messages in a group chat. Despite belonging to the same category of networks, further investigation \blue{into} the differences in their structural properties \blue{as they }evolv\blue{e over} time is needed.

\begin{landscape}
\begin{table}[h!]
\begin{adjustwidth}{-3cm}{-0cm}
  \centering
  \small
  \begin{tabular}{ c c c c c c c c c c c}
  \toprule
  
  \textbf{Data} & \textbf{Dataset} & $\mathbf{|V|}$ & \textbf{GCC} & $\mathbf{d}$ & \textbf{ASPL} & \textbf{PA} & $\mathbf{\gamma}$ & \textbf{Directed}  & \textbf{Add} & \textbf{Ref} \\ 
  \cdashline{1-11}
  
  \toprule

\multirow{2}{*}{\textbf{WWW}}\footnote{Node removal} & \makecell{(subset of)\\nd.edu} & $\sim 300k$ & 0.11 & $-$  & 7 & $\checkmark$& 2.1  & U & N &  \cite{barabasi_emergence_1999, Dorogovtsev_2002, adamic_1999, newman_random_2002}
 \\
& nd.edu & $2 \cdot 10^8$ & 
0.11 & $-$ & 16 &$\checkmark$& 2.1/2.7 & D & N & \cite{boccaletti_2006, Dorogovtsev_2002, szabo_clustering_2004, ravasz_hierarchical_2003, kong_experience_2008}
 \\

\cdashline{1-11}

\multirow{2}{*}{\textbf{Citation}} & Math1999 & $\sim 300k$ & 0.15 & $-$ & 8.46 & $\checkmark$ & 2.47 & U & E & \cite{boccaletti_2006, grossmann_2002, newman_coauthorship_2004, szabo_clustering_2004} 
 \\
& hep-th & $\sim 30k$ & \makecell{[0.33, 0.48] $\Uparrow$} & \makecell{[2E–04, \\ 1E–03] $\Downarrow$} & 6.91 & $\checkmark$ & [2, 3] & $-$ & E & \cite{santoro_time_varying_2011} 
 \\

\cdashline{1-11}

\multirow{3}{*}{\textbf{Co-authorship}} & Medline & 1.5M & [0.06, 0.15] $\Downarrow$ & $-$ & 4.6 & $\checkmark$ & $\checkmark$ &  $-$ & E & \cite{newman_pref_2001, newman_random_2002, Dorogovtsev_2002} \\

& Math 1991-98 & $\sim 70k$ & [0.85, 0.8] $\Downarrow$ & $\Uparrow$ & [16, 10] $\Downarrow$ & $\checkmark$ & 2.4 & $-$ & E & \cite{Dorogovtsev_2002, barabasi_pa_2001, barabasi_evolution_2002, Lee_2006} \\

& \makecell{Neuro-science\\1991-98} & $\sim 200k$ & [0.7, 0.6] $\Downarrow$ & $\Uparrow$ & [10, 6] $\Downarrow$ & $\checkmark$ & 2.1 & $-$ & E & \cite{Dorogovtsev_2002, barabasi_pa_2001, barabasi_evolution_2002, Lee_2006} \\

\cdashline{1-11}

\textbf{\makecell{Actor\\collaboration}} & IMDB & $\sim 200k$ & 0.79 & $-$ & 3.65 $\Downarrow$ & $\checkmark$ &  2.3 &  U & $-$ &  \makecell{\cite{barabasi_emergence_1999, boccaletti_2006, Dorogovtsev_2002} \\ \cite{ watts_collective_1998, newman_random_2002, ravasz_hierarchical_2003}}
 \\

\cdashline{1-11}

\multirow{2}{*}{\textbf{E-mail}} & DNC-email & $\sim 2k$ & $6.29$ & 0.021 & $-$ & $\checkmark$ & $-$ &  D & $-$ & \cite{kejriwal_2022, network_repo} \\

& Email-EU & $\sim 32k$ & 0.11 & 1E–04 & $-$ & $\checkmark$ & $-$ &  D & $-$ & \cite{kejriwal_2022, network_repo} \\

\cdashline{1-11}

\multirow{7}{*}{\textbf{Reddit users}} & PushShift & 500k & 0.15 & [0.02, 0.1] & $-$ & $-$ & $-$ & U & N & \cite{survival_2020, makow_2017}  \\

& r/Geopolitics & [2, $\sim 1k$] 
& 0.12 $\Downarrow$$^a$& 0.11 $\Downarrow$$^a$& 2.32 $\Uparrow$$^a$& $-$ & $-$ & D & E & \cite{zhu2022reddit} \\

& r/War & [2, $\sim 300$] 
& 0.12 $\Downarrow$$^a$& 0.13 $\Downarrow$$^a$& 1.8 $\Uparrow$$^a$& $-$ & $-$ & D & E & \cite{zhu2022reddit} \\

& r/PinoyProgrammer & [2, $\sim 200$] 
& 0.15 $\Downarrow$$^a$& 0.19 $\Downarrow$$^a$& 1.5 $\Uparrow$$^a$& $-$ & $-$ & D & E & \cite{bwandowando_pinoy} \\

& r/Ukraine & [2, $\sim 3k$] 
& 0.09 $\Downarrow$$^a$ & 0.1 $\Downarrow$$^a$ & 1.7 $\Uparrow$$^a$& $-$ & $-$ & D & E & \cite{bwandowando_ukraine, pohl2023invasion, dvzubur2022semantic} \\

& r/Jokes & [2, $\sim 5k$] 
& 0.01 $\Downarrow$$^a$& 0.13 $\Downarrow$$^a$& 1.7 $\Uparrow$$^a$& $-$ & $-$ & D & E & \cite{bwandowando_jokes} \\

 & r/AITA & [2, $\sim 12k$] 
 & \textbf{1E–05}$^a$ & 1E–04$^a$ & \textbf{2.7 $\Uparrow$}$^a$ & $\checkmark$ & [1.8, 3.5] & D & E & \\

  \bottomrule
  \end{tabular}
  \\[\baselineskip]
  \caption{State of the art of growing real social networks. The last row represents our data. The table reports: number of vertices $|$V$|$, global clustering coefficient GCC, density $d$, average shortest path length ASPL, preferential attachment PA, exponent of the degree distribution $\gamma$, type of network (directed or undirected), type of growth (adding nodes or edges).
$|$V$|$ is reported as an interval [min, max] for datasets containing more networks. 
The two values of $\gamma$ represent, respectively, the in/out-degree exponents when the network is directed. The $^a$ indicates that the measure has been averaged over multiple networks. Symbols $\Uparrow$ and $\Downarrow$ are used to indicate that a metric, respectively, increases or decreases over time.}
  \label{tab:sota_growing_net}

\end{adjustwidth}
\end{table}
\end{landscape}

\section{Methodological framework}
\label{sec:method}

In order to study online conversations in which users express moral judgments, we collect data from the AITA community (Section \ref{subsec:data}) and operationalize the judgment behavior of participants (Section \ref{subsec:judgment}).
Then, we provide a measure for disagreement among users, \blue{representing} how much polarizing their judgments are (Section \ref{subsubsec:disagreement}). Finally, we model each conversation as a growing complex network and we study its evolution in time (Section \ref{subsec:temporal_network}).

\subsection{Data}
\label{subsec:data}

\begin{figure}[h]
  \centering

  \begin{subfigure}{0.49\linewidth}
    \centering
    \includegraphics[width=\linewidth]{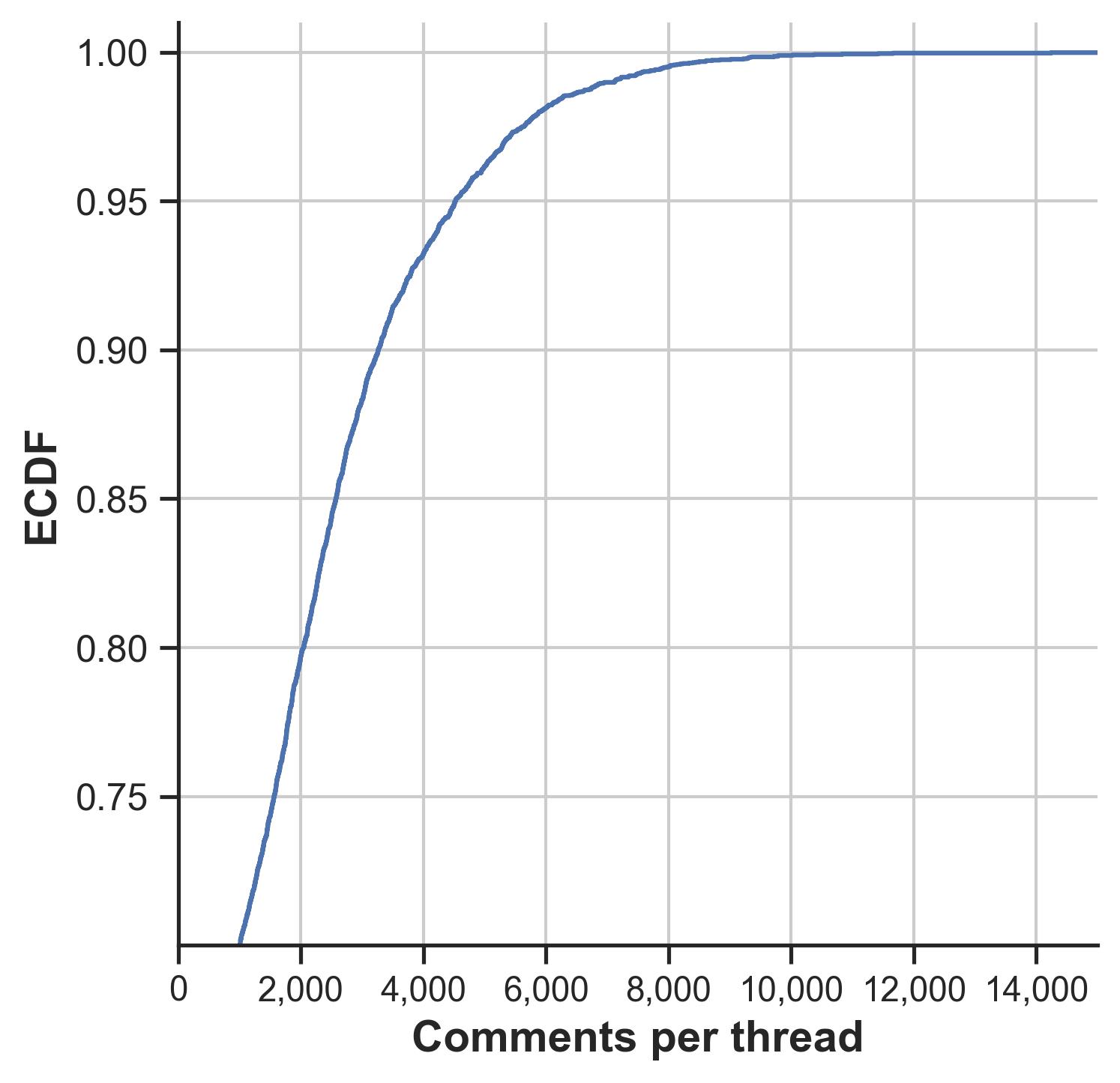}
    \caption{}
    \label{ecdf_thread_dim}
  \end{subfigure}
  \hfill
  \begin{subfigure}{0.49\linewidth}
    \centering
    \includegraphics[width=\linewidth]{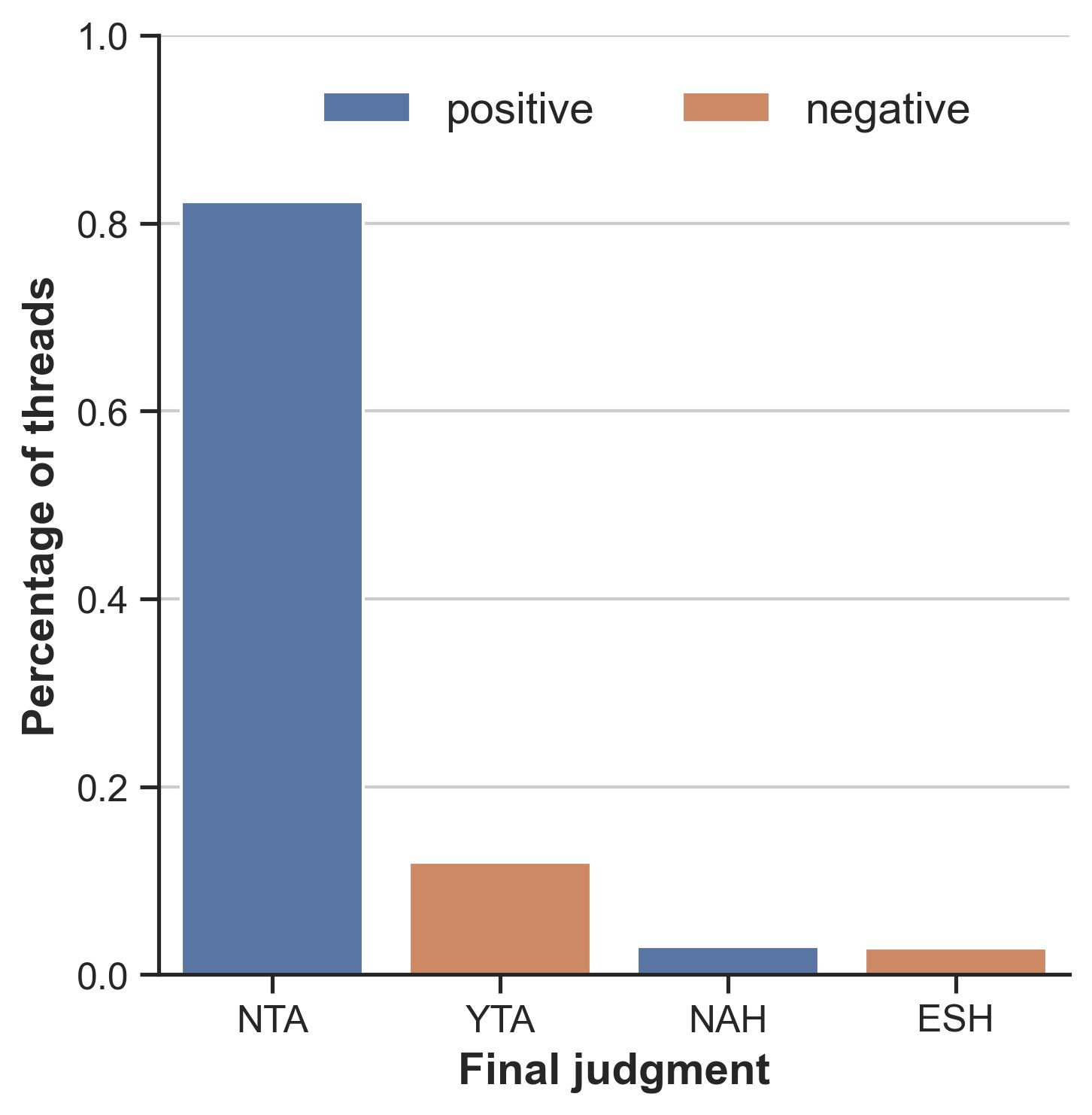}
    \caption{}
    \label{final_judg_distr}
  \end{subfigure}
  \caption{Statistics of the AITA threads dataset. \blue{In (a) the ECDF of the size (number of comments) per threads while in (b) distribution of final verdicts of the threads.}}
  \label{fig:comm_distrib}
\end{figure}

We downloaded 6,366 threads, containing \blue{a} total \blue{of} 6,372,251 comments, from the AITA subreddit using the PRAW library\footnote{Python Reddit API Wrapper (\url{https://praw.readthedocs.io/en/stable/})}. In particular, we download the ``top'' submissions \blue{---} those having the highest score\blue{,} measured as the difference between upvotes and downvotes of a post (i.e. the thread root). By definition, top posts are likely to have received significant attention, possibly resulting in a large volume of comments. \blue{In order to gather a representative dataset, we performed 10 different queries across various temporal scopes, ranging from one week to multiple years,} each gathering different sets of top submissions \blue{along} with all the comments.  We set the limit of each query to 1,000 to comply with the Reddit API limits \footnote{\url{https://support.reddithelp.com/hc/en-us/articles/16160319875092-Reddit-Data-API-Wiki}} and we removed duplicated threads. The final dataset size is reported in Table \ref{tab:subm}, which also contains the temporal scope of data selection. 
Figure \ref{ecdf_thread_dim} shows the distribution of thread \blue{sizes} (measured as the number of comments), while Figure \ref{final_judg_distr} shows the distribution of final verdicts across threads. Note that 75\% of the threads have less than 2,000 comments\blue{,} and 80\% of them have been assigned ``NTA'' as final verdict.

\begin{table}[h!]
  \centering
  \small
  \begin{tabular}{ c c c c }
  \toprule
  
  \textbf{Query} & \textbf{Temporal scope} & \textbf{Threads} & \textbf{Comments} \\ \cdashline{1-4}
  
  \toprule

        1 & Week 38, 2023 & 862 & 149,834 \\ \cdashline{1-4}
        2 & Week \blue{39}, 2023 & 883 & 149,814 \\ \cdashline{1-4}
        3 & Week 40, 2023 & 560 & 81,006 \\ \cdashline{1-4}
        4 & Week 41, 2023 & 881 & 139,643 \\ \cdashline{1-4}
        5 & Week 42, 2023 & 881 & 172,033 \\ \cdashline{1-4}
        \blue{6} & Month (Oct 2023) & 526 & 349,540 \\ \cdashline{1-4}
        \blue{7} & Year (2023) & 931 & 2,998,481 \\ \cdashline{1-4}
        \blue{8} & Year (2023) & 91 & 258,627 \\ \cdashline{1-4}
        \blue{9} & All time & 740 & 2,038,234 \\ \cdashline{1-4}
        \blue{10} & All time & 11 & 35,039 \\ \hline
            Total & & 6,366 & 6,372,251 \\
  \bottomrule
  \end{tabular}
  \\[\baselineskip]
  \caption{Data collection of top submissions from the AITA subreddit. We ran a total of ten queries during September and October 2023, collecting the top threads with different temporal scopes. \blue{The columns represent the query number, the specific time period covered by each query, the number of threads retrieved, and the total number of comments in those threads.}}
  \label{tab:subm}

\end{table}



\subsection{Operationalization}
\label{subsec:operationalization}

\subsubsection{Judgment behavior.}
\label{subsubsec:judgment}

In the AITA community\blue{,} users \blue{participate} by writing posts (to be judged by others) or comments (to judge others). This paradigm established by the community implies that people commenting are expected to express a vote. We distinguish between \textit{voting} (i.e. writing comments containing at least one acronym among those listed in Table \ref{tab:flair_acronyms}) and \textit{discussing} (writing text without expressing a vote). For the purpose\blue{s} of this work, we decided to disregard the INFO acronym\blue{, as} it \blue{does} not \blue{constitute} a vote by definition.

The AITA community has specific guidelines about how users should vote and how the votes are processed to obtain the final verdict. Users can access these rules from the dedicated page\footnote{\url{https://www.reddit.com/r/AmItheAsshole/about/rules}}, the FAQ page\footnote{\url{https://www.reddit.com/r/AmItheAsshole/wiki/faq/}}, or the ``Voting rules'' section in the navigation panel of the homepage. Th\blue{e}se resources are also \blue{referenced in} every post since a bot automatically includes them in a top-level comment produced as soon as the post is published. Such comment is pinned on top \blue{for} maximum visibility, so users are aware of how they are expected to behave. According to the AITA rules, users must vote including one and only one voting label in their top-level comment. This implies that: (i) users cannot include more than one label in the text, (ii) the label should be one of those provided by the community \blue{and} correctly spelled\blue{, and} (iii) the comment containing it must appear in the first level of the thread. The label can appear at any point in the text and does not necessarily have to be capitalized. Since the judgment process (votes and upvotes) lasts 18 hours, the comments should also be published \blue{within} this time window to be part of the voting contest.

\subsubsection{Disagreement.}
\label{subsubsec:disagreement}

\blue{T}o measure disagreement of AITA threads, we \blue{use} the codified information about judgments expressed by users. As mentioned \blue{earlier}, in the AITA community, users explicitly tak\blue{e} a side and mak\blue{e} it public \blue{when they express a vote}. \blue{Consequently}, we label each comment with the respective judgment label. The voting labels represent the sides that users are taking, \blue{making} it straightforward to \blue{determine} which side each comment belongs \blue{to}. In this context, we measure \blue{the level of} disagreement in a thread by measuring the uncertainty of the judgments expressed in the comments. We do this by computing the probability of each label appearing (i.e., of each side to be taken) and measuring the Shannon entropy of the post.

Following~\cite{de_candia_social_2022}, who used binary entropy on aggregated votes to measure controversiality, we use multi-label entropy to operationalize disagreement. 

\begin{figure}[h]
    \centering
    \includegraphics[width=1\textwidth]{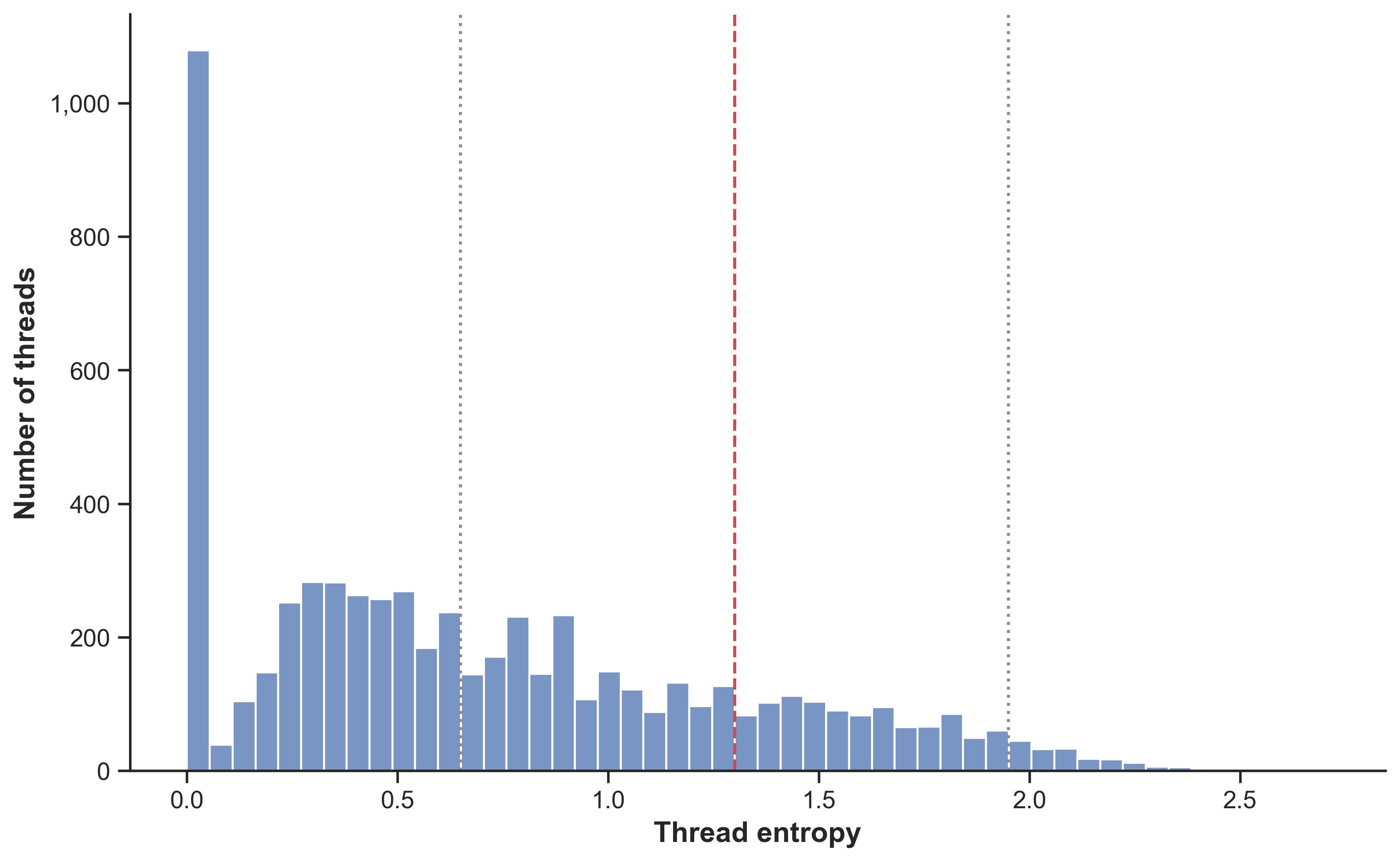}
    \caption{Distribution of the disagreement (computed as a thread entropy) across all the collected threads from the AITA community. Values fall in the range [0, 2.6]. Vertical lines divide the disagreement in low ($H < 0.65$), medium-low ($.65< H < 1.3$), medium-high ($1.3< H < 1.95$) and high ($H > 1.95$).}
    \label{fig:entropy}
\end{figure}

Given a the set of labels $\mathcal{X}$, the entropy of a post is defined as: 
\begin{equation}\label{eq:entropy}
    H(X) = - \sum_{x \in \mathcal{X}} p(x) \log p(x)
\end{equation}
where $p(x)$ is the discrete probability distribution of the labels appearing in the comments of the post.
Since we do not consider the INFO label, we have six possible labels (see Table \ref{tab:flair_acronyms})\blue{, so} the maximum value of entropy for each post is $\log_2 |X| \approx 2.6$.
Values of the entropy close to 2.6 indicate maximum uncertainty and therefore maximum divisiveness: judgments are uniformly split among the different labels, \blue{with} people equally tak\blue{ing} all the different sides. In this case, we can say that the post has high disagreement.
In contrast, a value of 0 would represent the maximum level of certainty: all judgments are unanimous and users all agree on taking one side\blue{,} so the post has no disagreement.
As shown in Figure \ref{fig:entropy}, around 53\% of the posts have low entropy ($< 0.65$)\blue{, indicating that} in more than half of the posts people agree on the judgment. 

\subsection{Temporal network analysis}
\label{subsec:temporal_network}

\begin{figure}[t!]
    \centering
    \includegraphics[width=\textwidth]{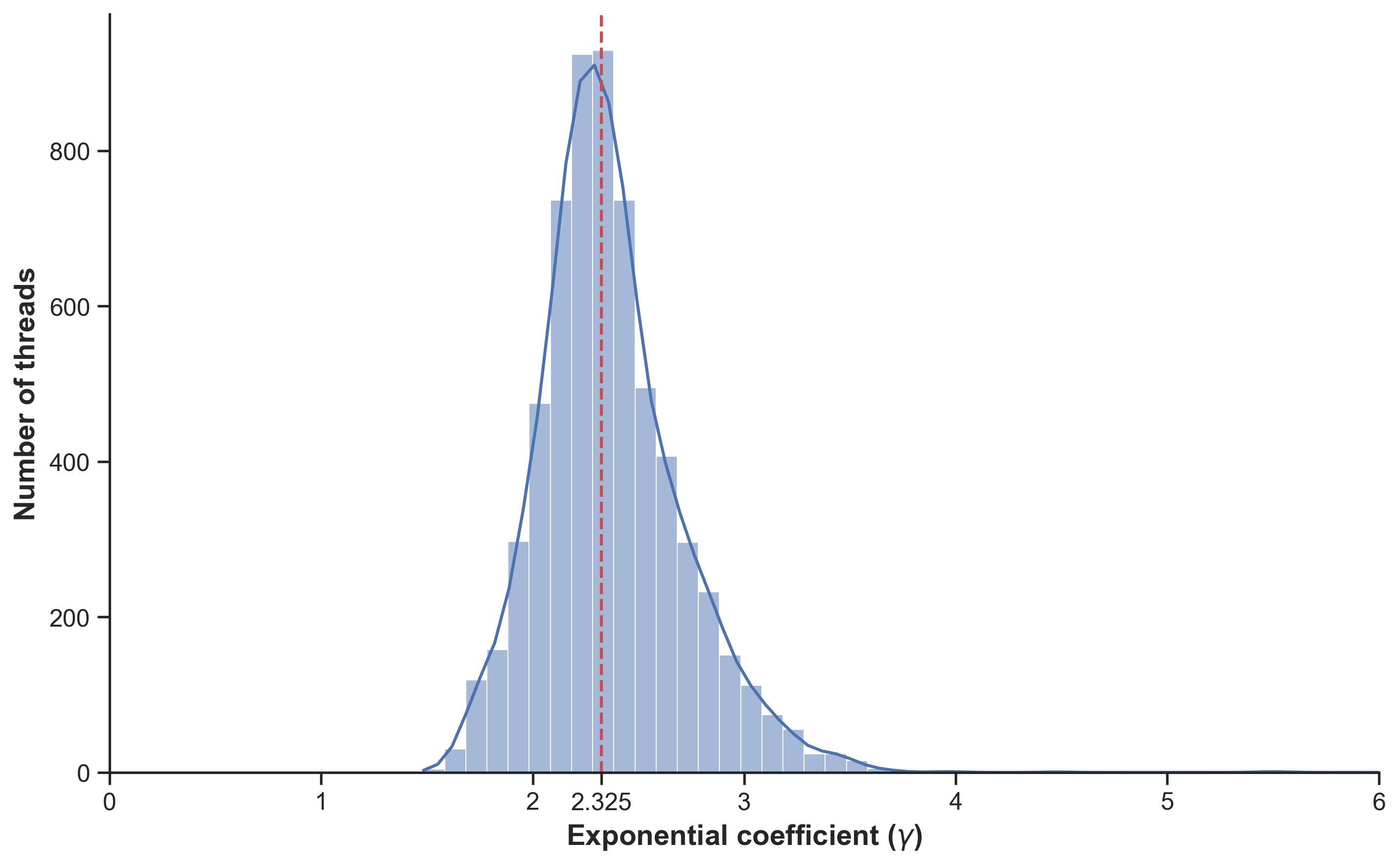}
    \caption{Distribution of coefficients ($\gamma$) of AITA networks' degree distribution.}
    \label{fig:scale_free}
\end{figure}

\blue{We model the discussions collected from the AITA community as networks of user interactions. For each thread, we build} a directed multi-graph $\blue{M} = (V, E, \blue{t, x})$ with attributed nodes and edges. The set of vertices $V$ represents users and the set of edges $E$ represents the answering comments. 
We extract the voting acronyms of each comment and we store them as a \blue{vertex} attribute set $X$. \blue{Hence, $x: V \rightarrow X$ is a function assigning to each vertex, the set of judgments expressed by that user in their comments.}
Since we could not determine the expressed vote from comments containing different acronyms (e.g., [``NTA", ``ESH", ``YTA"]) we 
label those judgments as \textit{unsure}\footnote{\blue{Note that we did not include ``unsure'' comments in the disagreement computation described in the previous section, since it is impossible to infer from such comments which judgment users are willing to express.}}.
The temporal information is \blue{embedded} by a scalar $t: E \rightarrow T$, stored as an edge attribute, where $T$ is an ordered set of time annotations \blue{with a resolution of seconds}.
We perform a statistical test to prove that such networks are scale-free. This because, in order to compare our network with the state of the art on real social networks, we first need to demonstrate that our networks are scale-free, i.e. that their degree distribution follows a power law distribution $k^\gamma$, where $2 < \gamma < 3$. Hence we fit our empirical data to a power-law distribution 
and we measure the distribution of the exponents to verify that they mostly fall in the range [2, 3]. The results are shown in Figure \ref{fig:scale_free}. \blue{To assess the goodness of the fit we performed a one-sample Kolmogorov-Smirnov (KS) test for all the degree distributions of the networks, which returned a coefficient smaller than .35 for all the networks and a p-value greater than .001 for 88\% of the networks, confirming that the empirical distribution of our data is (significantly) very close to a power-law distribution.}
\blue{Such results} confirm that AITA networks are scale-free, hence we can compare their properties with other real social networks.

Then\blue{,} we study \blue{each} network \blue{$M$} of user interactions from a temporal perspective, by reconstructing them in time. We obtain, for each thread, a \blue{set of }directed network\blue{s} \blue{$G = \langle G_1, ..., G_k \rangle$} that grow over time\blue{, where each network $G_k = (V_k, E_k), k = 0\dots|E|$ is the k-\emph{th} network. Therefore, $V_k \subseteq V$ includes the user starting the thread and all users commenting until k-\emph{th} messages have been posted. Each edge in the set $E_k = (v, u) \subseteq E$ indicates that user $v$ has written at least one comment to user $u$.}


\section{Results}
\label{sec:results}
 
\begin{figure}[h!]
    \centering
    \includegraphics[width=\textwidth]{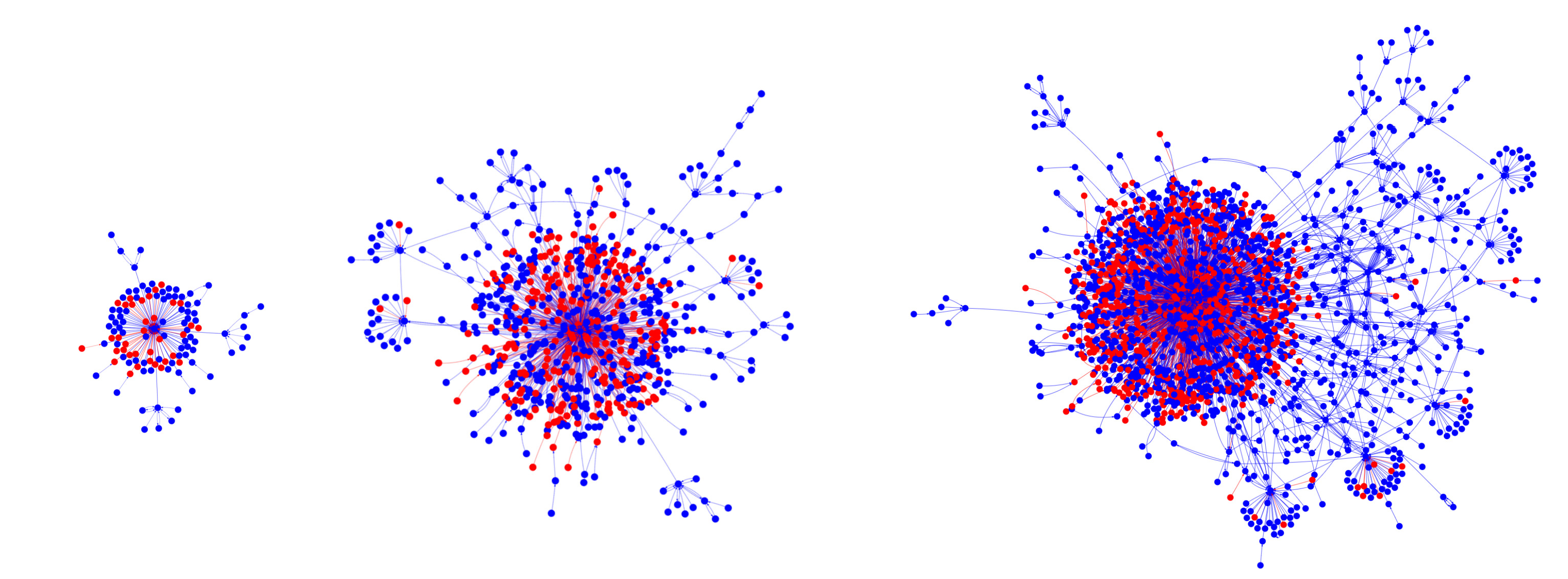}
    \caption{An example of user interaction network built from an AITA thread. Nodes (voters in red, not voters in blue) are users and direct edges $e_{ij}$ represent comments from user $i$ to user $j$. The graph shows two different sub-structures: a star, with the hub node corresponding to the original poster and everyone who replied to it, and a periphery consisting of comments and replies among users.}
    \label{fig:growing_net}
\end{figure}

\begin{figure}[h!]
    \centering
    \includegraphics[width=0.7\textwidth]{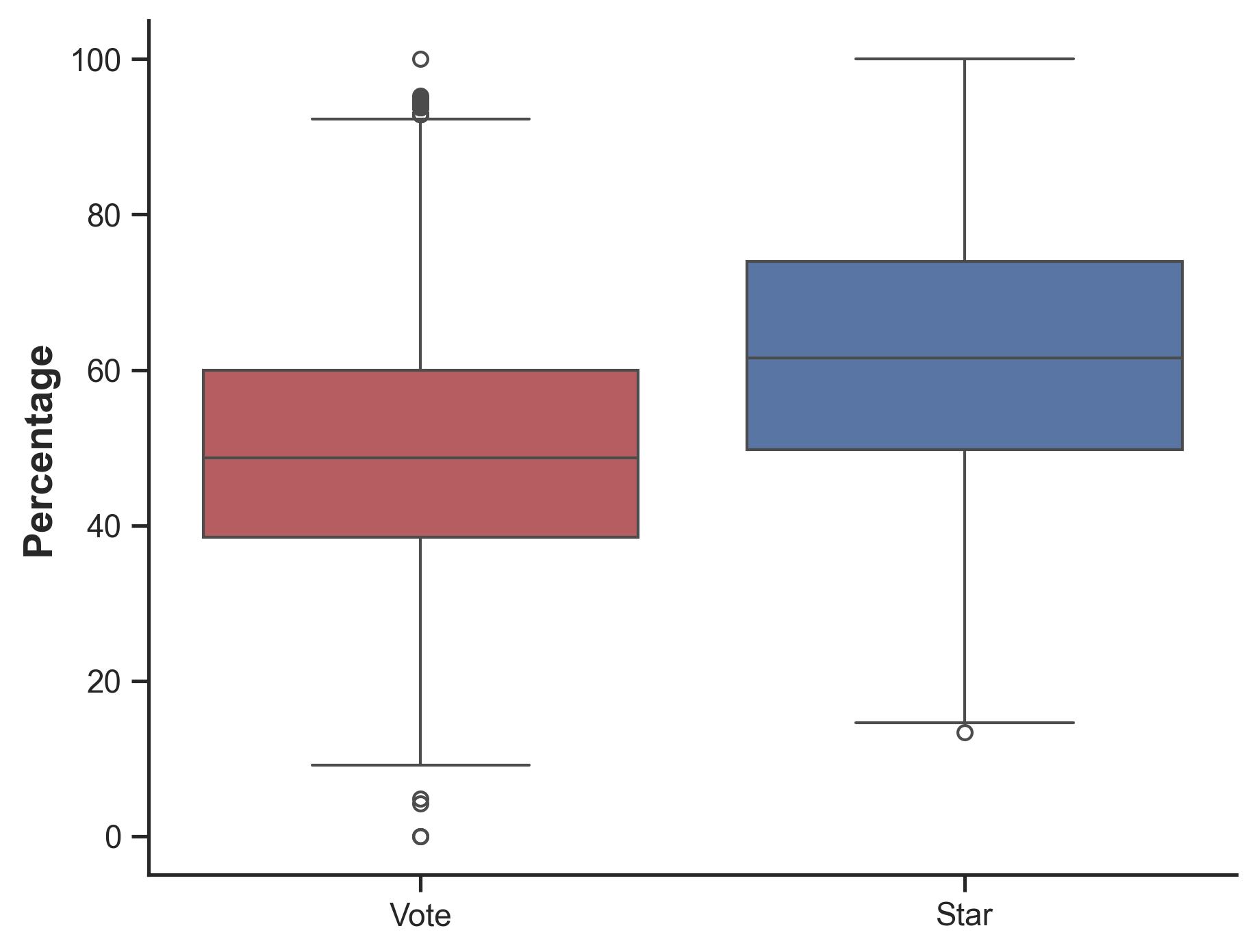}
    \caption{Distribution of users expressing their opinion as a vote (left) and of users joining the thread with a first-level comment directly to the author of the post (right). \blue{On average, (i) half of participants express a vote, and (ii) 60\% of users in the AITA community join the thread in the star. Conversely, the periphery includes most of the participants discussing without expressing a vote.}}
    \label{fig:user_distribution}
\end{figure}

\blue{As} users join the conversation \blue{thread}, \blue{new} interactions \blue{are} form\blue{ed over time,} and the network grows\blue{. T}he dynamic evolution of the network generates two \blue{distinct} subgraphs: one \blue{consisting of} participants directly \blue{responding} to the author of the post \blue{(i.e., users writing first-level comments)}, and the \blue{other comprising} users joining with comments located at deeper levels in the thread. We refer to these subgraphs as the \textit{star} and the \textit{periphery}, respectively. Figure~\ref{fig:growing_net} \blue{illustrates} one of the AITA networks evolving \blue{over} time\blue{, demonstrating} how these interactions and subgraphs develop. Red nodes represent users voting in at least one comment, while blue nodes represent users writing comments without expressing a vote (i.e.\blue{,} discussing).
\blue{Figure \ref{fig:user_distribution} shows that most of the voters are located in the star, a consequence of the community rules, which state that votes should be expressed in first-level comments. We describe in detail how this rule impacts user behavior in the community in Section \ref{sec:conclusions}.}

\blue{In the following subsections, we provide different views of these networks of interactions and their subgraphs, and investigate whether the guidelines of the AITA subreddit would result in significantly different structural and growing properties. First, we describe \textit{why} the star and the periphery exist, and we explore the response time of comments in the network (Section \ref{subsec:results_substructures}).} In Section \ref{subsec:results_aita} we analyze \blue{\textit{how} the networks growth from a global perspective, by} compar\blue{ing the evolution of their structural properties} with the state of the art of real dynamic social networks, \blue{previously} summarized in Table \ref{tab:sota_growing_net}. 

\blue{Then, in Section \ref{subsec:results_reddit} we compare the growth of the two substructures of AITA networks with networks from other subreddits, \blue{concluding} that the \blue{growth speed} of the star \blue{is between 2 and 3 times faster} than \blue{that of the} periphery subgraph\blue{. This difference} is significantly larger than in other subreddits. Finally, in Section \ref{subsec:reciprocity} we \blue{examine} the relation between thread entropy and other features of the threads to demonstrate that disagreement plays a role in the discussions of the AITA community.}

\subsection{\blue{Response time}}
\label{subsec:results_substructures}

\blue{To capture how quickly users participate in the star and in the periphery, we compute how fast they respond to a message (i.e., how fast a replying edge is added in each subgraph). We calculate the time differences between a comment and its parent node (the post-root or the preceding comment) and we refer to this quantity as the response time $R$. We only consider response times within the range of $[\mu - 2\sigma, \mu + 2\sigma]$ to exclude outliers, where the $\mu$ is the mean response time of parent-child edges in the given graph and $\sigma$ represents one standard deviation from the mean.

Figure \ref{fig:response_time} shows the distribution of response times in the star (left) and in the periphery (right), both for comments containing a vote (blue) or not (red). On average, the response time in AITA threads is between $10^4$ and $10^5$ seconds. Our main interest lies in the periphery, where we observe that the response time of voting comments is higher than non-voting comments, suggesting that writing a comment that contains a judgment requires more time. In Section \ref{sec:conclusions} we discuss this phenomenon in depth in relation with the AITA community guidelines and with SJT. The difference in response time between voting and non-voting comments in the star is neither interesting, due to a large imbalance in the data --- with more than 70\% of voting comments in the star ---, or statistically significant.

\begin{figure}[h]
\centering
\includegraphics[width=0.8\textwidth]{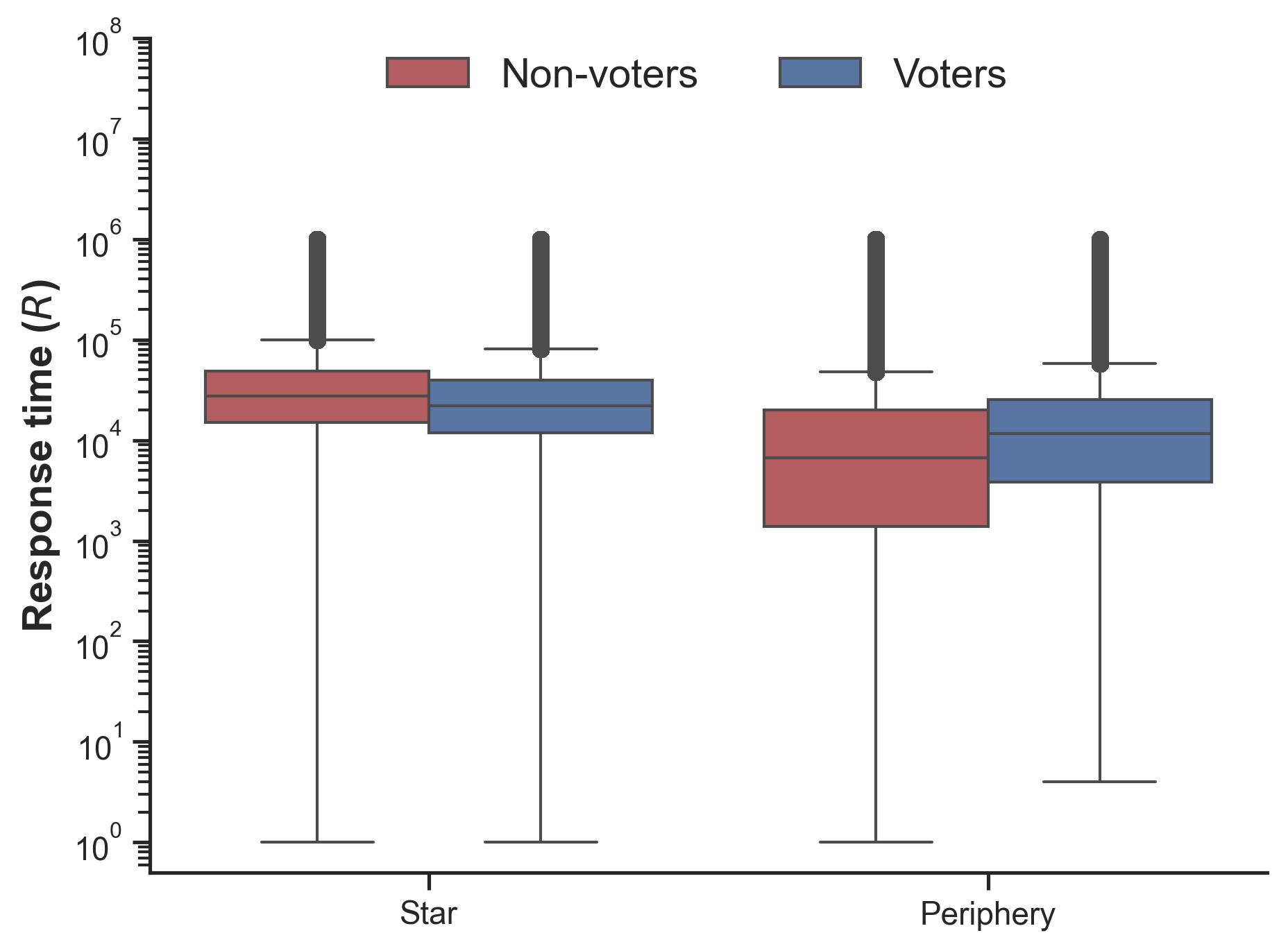}
    \caption{\blue{Response time of AITA threads. In the star, the  $\bar{R}$ of voting comments is $\sim 3$ hours while it is $\sim 2$ hours for non-voting comments. In the periphery, the $\bar{R}$ is $\sim 6$ hours for voting comments and $\sim 7.5$ hours for non-voting comments. }}
    \label{fig:response_time}
\end{figure}

Finally, note that the difference in the average response time between the star and the periphery is very small, and is likely an artifact of how the measure has been constructed. While the response time in the periphery always represents the difference between a comment and the immediate reply, that is not the case for the star. In the star subgraph the response time will always increase as the networks grows since the parent comment is the root (post). For instance, the $R$ between a given comment and the root will always be larger than the distance between a previous comment and the root.}

\subsection{\blue{Structural properties of AITA evolving over time}}\label{subsec:results_aita}

According to the literature, the average shortest path length of growing real social networks usually decreases over time~\cite{leskovec_graphs_2005, barabasi_pa_2001, barabasi_evolution_2002, Lee_2006, barabasi_emergence_1999, boccaletti_2006, Dorogovtsev_2002, watts_collective_1998, newman_random_2002, ravasz_hierarchical_2003}. This happens \blue{because} the average number of steps needed to connect two random individuals tends to become relatively small due to the increasing number of paths available. 
\blue{As the} network \blue{expands} over time, more connections are established\blue{, increasing} the likelihood of finding shorter paths between individuals. The literature attributes this phenomenon to (i) the presence of highly connected individuals (``hubs”) that reduce the distance between different parts of the network, and (ii) the tendency for networks to exhibit a clustered structure, creat\blue{ing} local neighborhoods or communities within the network. Hence, real social networks that are scale-free exhibit preferential attachment and community structure, both contributing to shortening the average path length~\cite{barabasi_emergence_1999, PATTANAYAK20228401, sallaberry_model_2013}.

\begin{figure}[h!]
  \centering

  \begin{subfigure}{0.49\textwidth}
    \centering    \includegraphics[width=\textwidth]{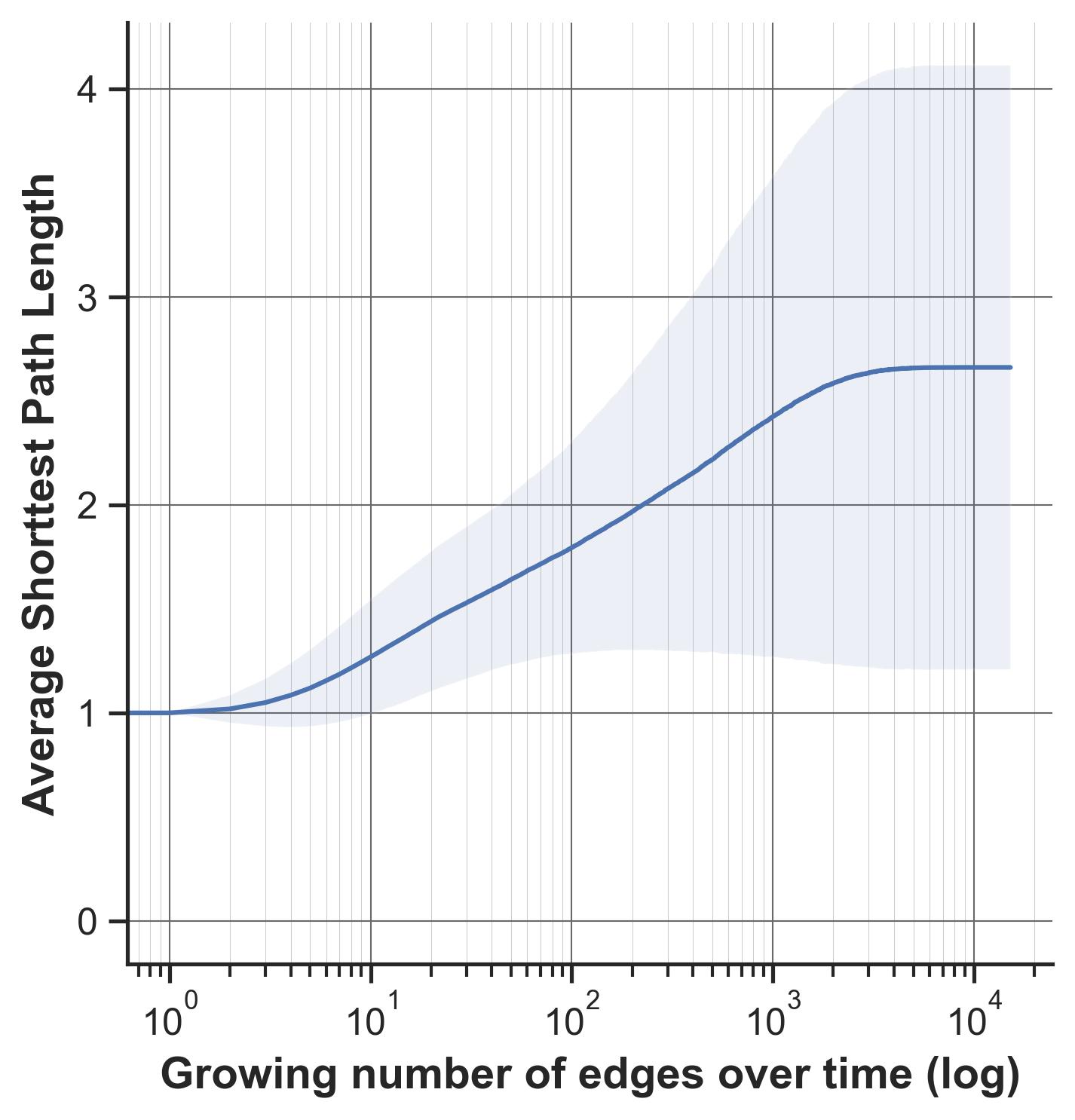}
    \caption{ASPL}
    \label{subfig:aspl_in_time}
  \end{subfigure}
  \hfill
  \begin{subfigure}{0.49\textwidth}
    \centering
    \includegraphics[width=\textwidth]{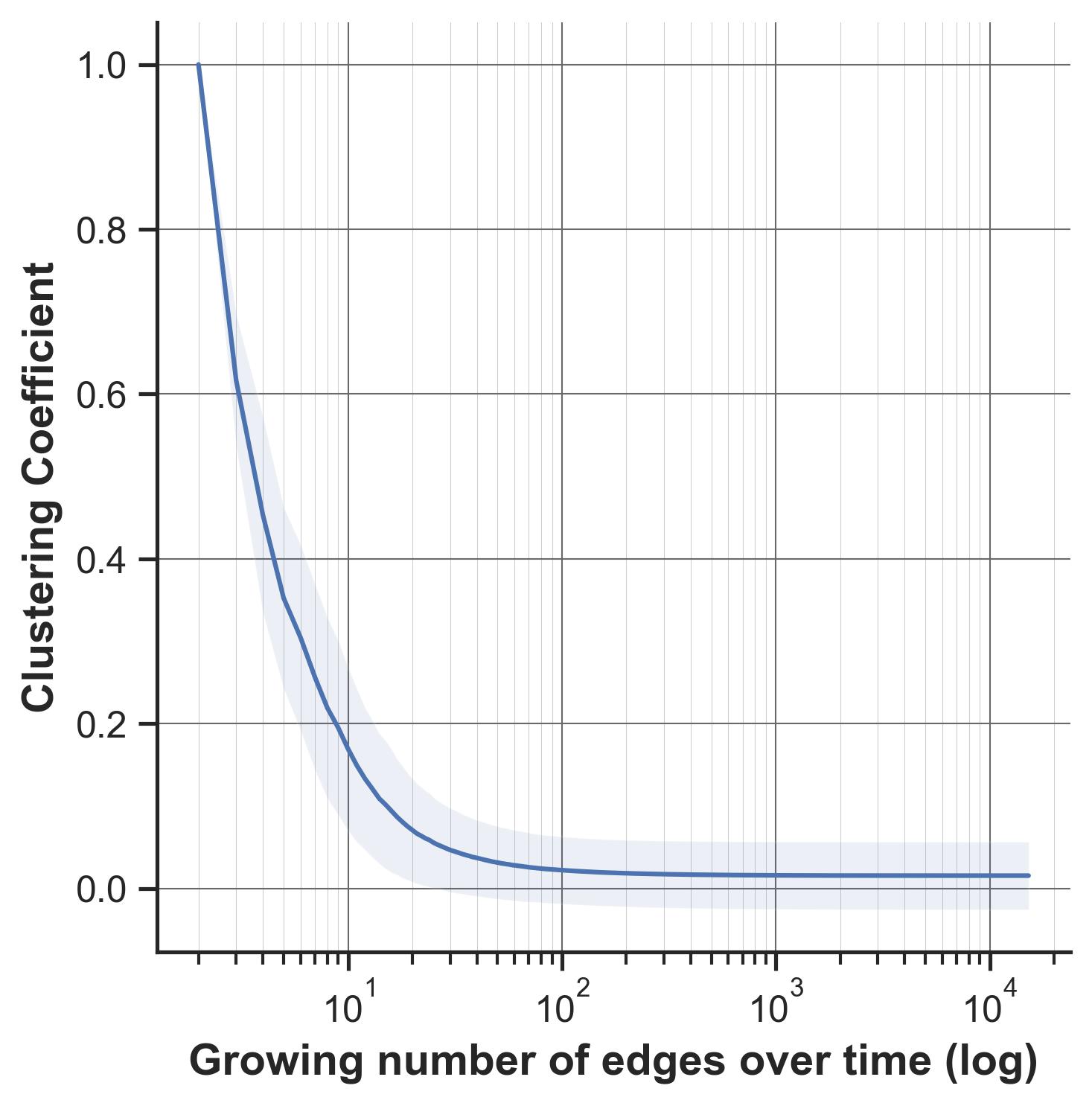}
    \caption{GCC}
    \label{subfig:clust_coeff_in_time}
  \end{subfigure}

  \medskip

  \begin{subfigure}{0.49\textwidth}
    \centering
    \includegraphics[width=\textwidth]{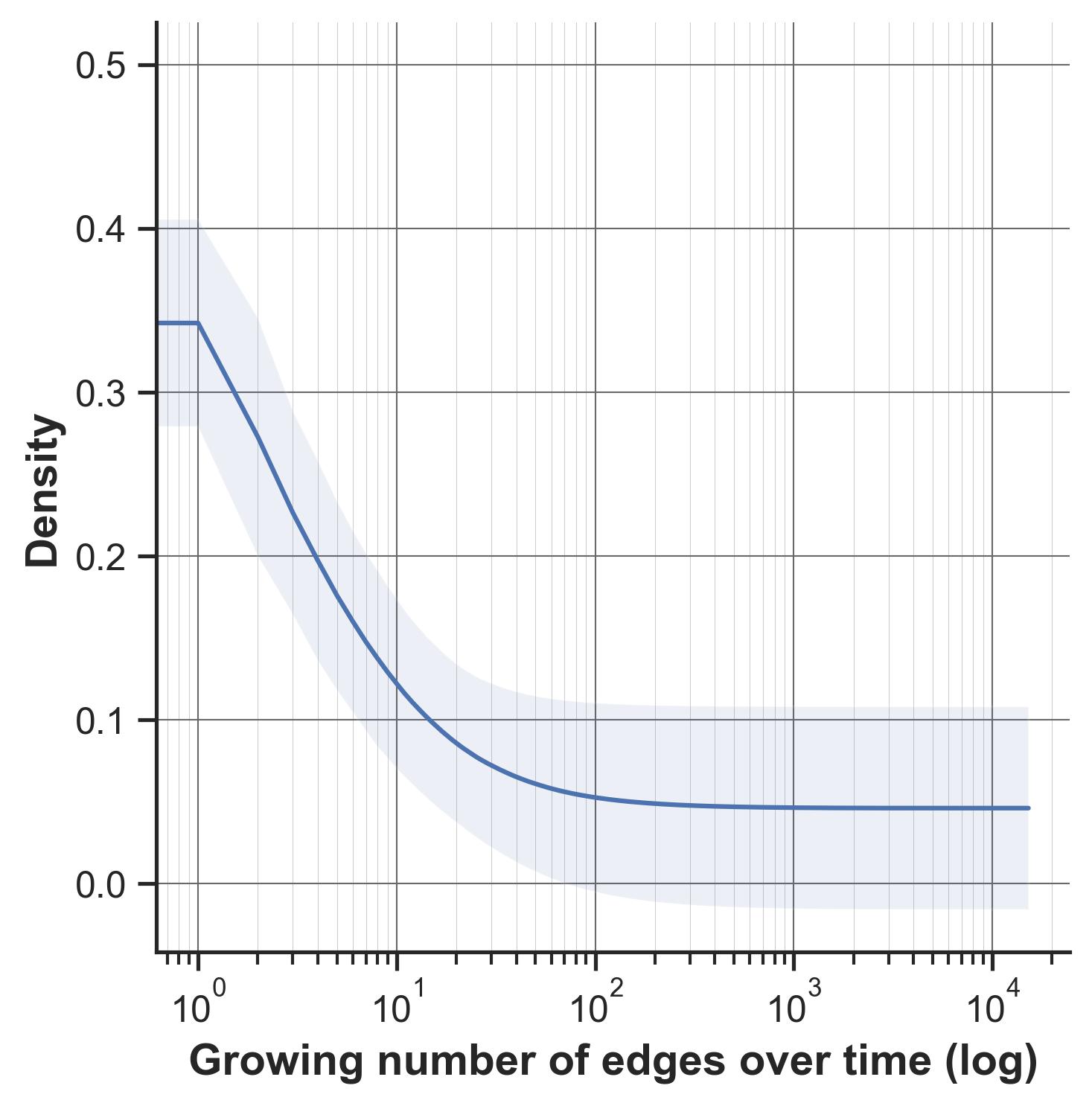}
    \caption{$d$}
    \label{subfig:density_in_time}
  \end{subfigure}
  \hfill
  \begin{subfigure}{0.49\textwidth}
    \centering
    \includegraphics[width=\textwidth]{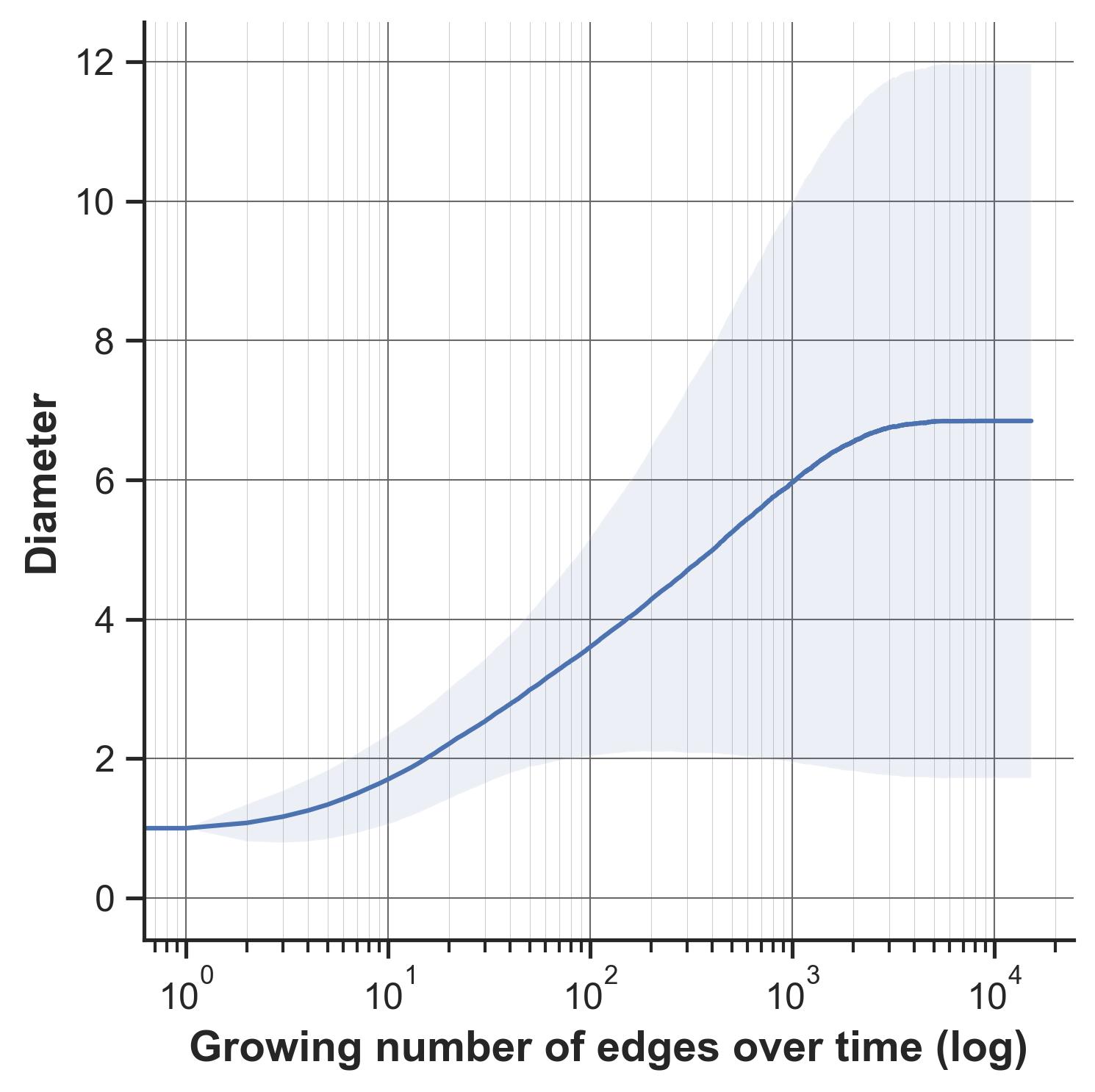}
    \caption{D}
    \label{subfig:diameter_in_time}
  \end{subfigure}

  \caption{Structural properties of AITA networks growing \blue{over} time. Each plot shows the metric averaged over all the networks at each timestamp when a new edge is added: \blue{(a) average shortest path length (ASPL), (b) global clustering coefficient (GCC) (c) density (\textit{d}), and (d) diameter (D).}}
  \label{fig:metrics_in_time}
\end{figure}

In this work, we \blue{demonstrate} that Reddit networks of user interactions evolve differently \blue{from} what is described in the literature about growing real social networks. \blue{Specifically, during the network reconstruction process \blue{(explained in Section \ref{subsec:temporal_network})}, every time a new edge is added, we calculate the following structural properties of the network: density ($d$), global clustering coefficient (GCC), average shortest path length (ASPL) and diameter (D).}
We show that despite being scale-free (see Section \ref{subsec:temporal_network}), their ASPL increases with time. Moreover, their \blue{global clustering coefficient (}GCC\blue{)} is five orders of magnitude smaller than expected since, on average, an extremely small number of clusters \blue{are} formed. \blue{Figure \ref{fig:metrics_in_time} shows the evolution of these metrics \blue{over} time for all threads (i.e.\blue{,} averaging the metric value at each timestamp over all the networks). The more edges are created over time, the more the ASPL increases while the GCC decreases. Moreover, the GCC is, on average, very small.}

This unexpected behavior of the network is what causes the increase of ASPL over time. Table \ref{tab:sota_growing_net} shows the state of the art of real social networks growing over time. Note that all the examples contained in the table have a high GCC and, when available, a decreasing ASPL over time.
By comparing the last row, which represents our AITA networks, with other rows\blue{,} it is clear that the GCC is negligible and that the ASPL behaves differently when such networks evolve: the more edges are added, the more \blue{the ASPL} increases over time.

\subsection{\blue{Growing networks of Reddit threads}}
\label{subsec:results_reddit}

\blue{In this section, we examine the growth speed of the two substructures in the AITA subreddit and compare it with other subreddit networks where the community rules do not incentive a particular behavior. For our comparison, we use five distinct pre-existing subreddits, which are openly available online and include temporal information of the comments. W}e pre-process \blue{these datasets by} removing threads containing \blue{fewer} than 2 comments, as well as \blue{duplicate }comments. Table~\ref{tab:new_datasets_stats} shows basic statistics of the data\blue{sets used} after \blue{pre-processing them}. \blue{For each dataset, w}e reconstruct \blue{its} conversation\blue{s} \blue{over} time following the same methodology described in Section \blue{\ref{subsec:temporal_network}}.

\begin{figure}[p]
    \centering
    \begin{subfigure}[b]{0.45\textwidth}
        \centering
        \includegraphics[width=\textwidth]{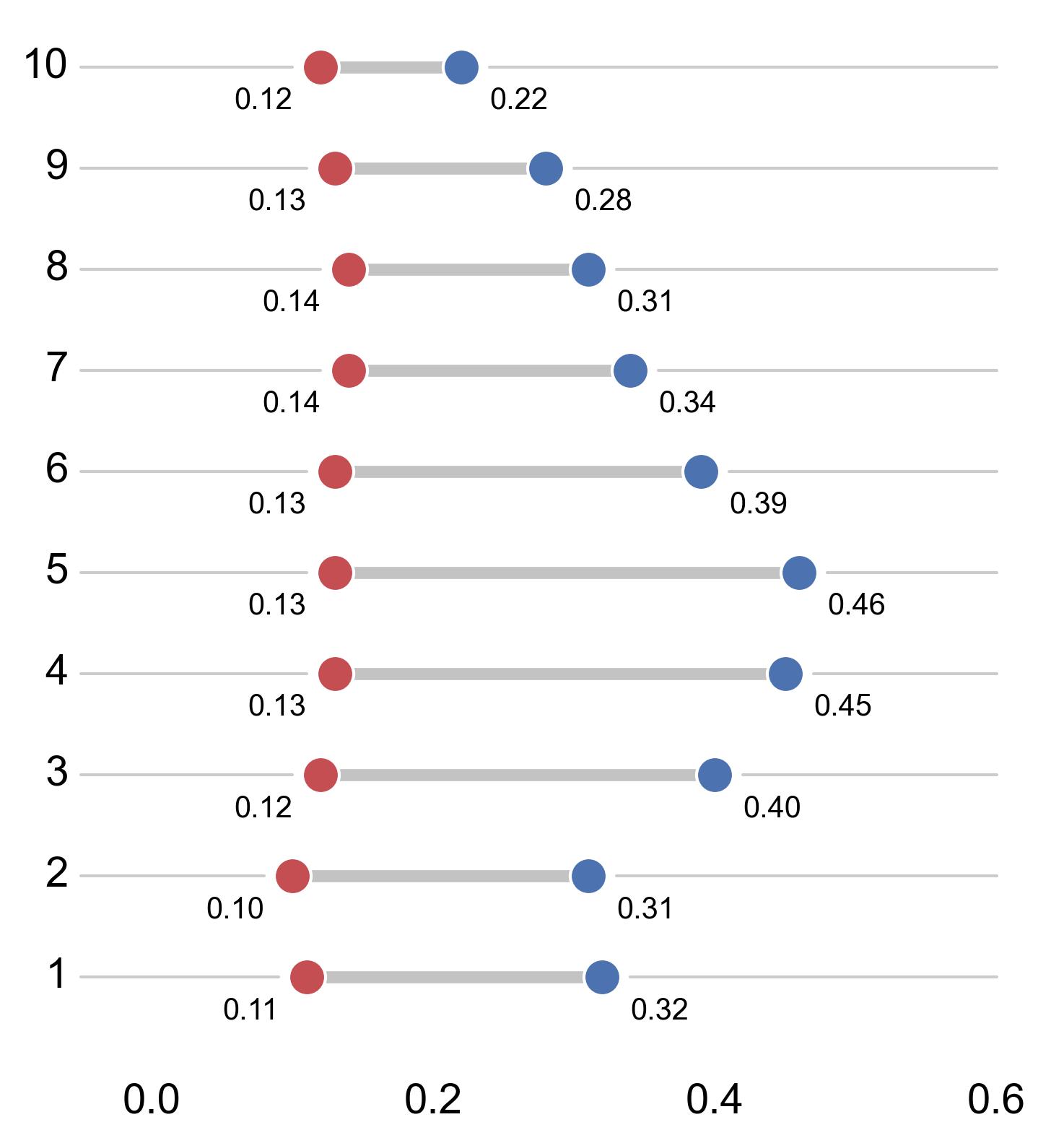}
        \caption{r/AITA}
        \label{fig:AITA}
    \end{subfigure}
    \hfill
    \begin{subfigure}[b]{0.45\textwidth}
        \centering
        \includegraphics[width=\textwidth]{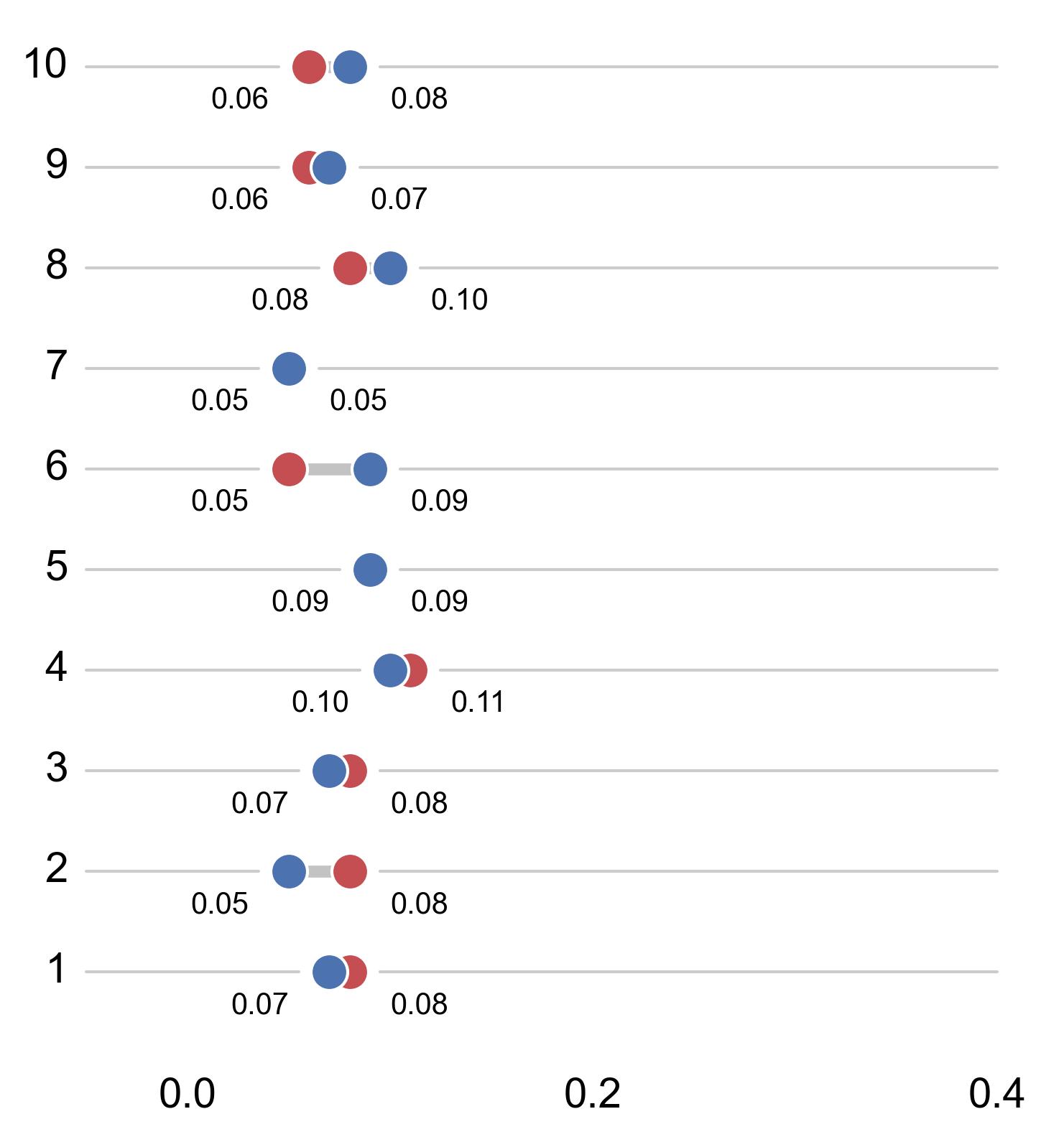}
        \caption{r/Geopolitics}
        \label{fig:Geopolitics}
    \end{subfigure}
    \vspace{0.5cm}
    
    \begin{subfigure}[b]{0.45\textwidth}
        \centering
        \includegraphics[width=\textwidth]{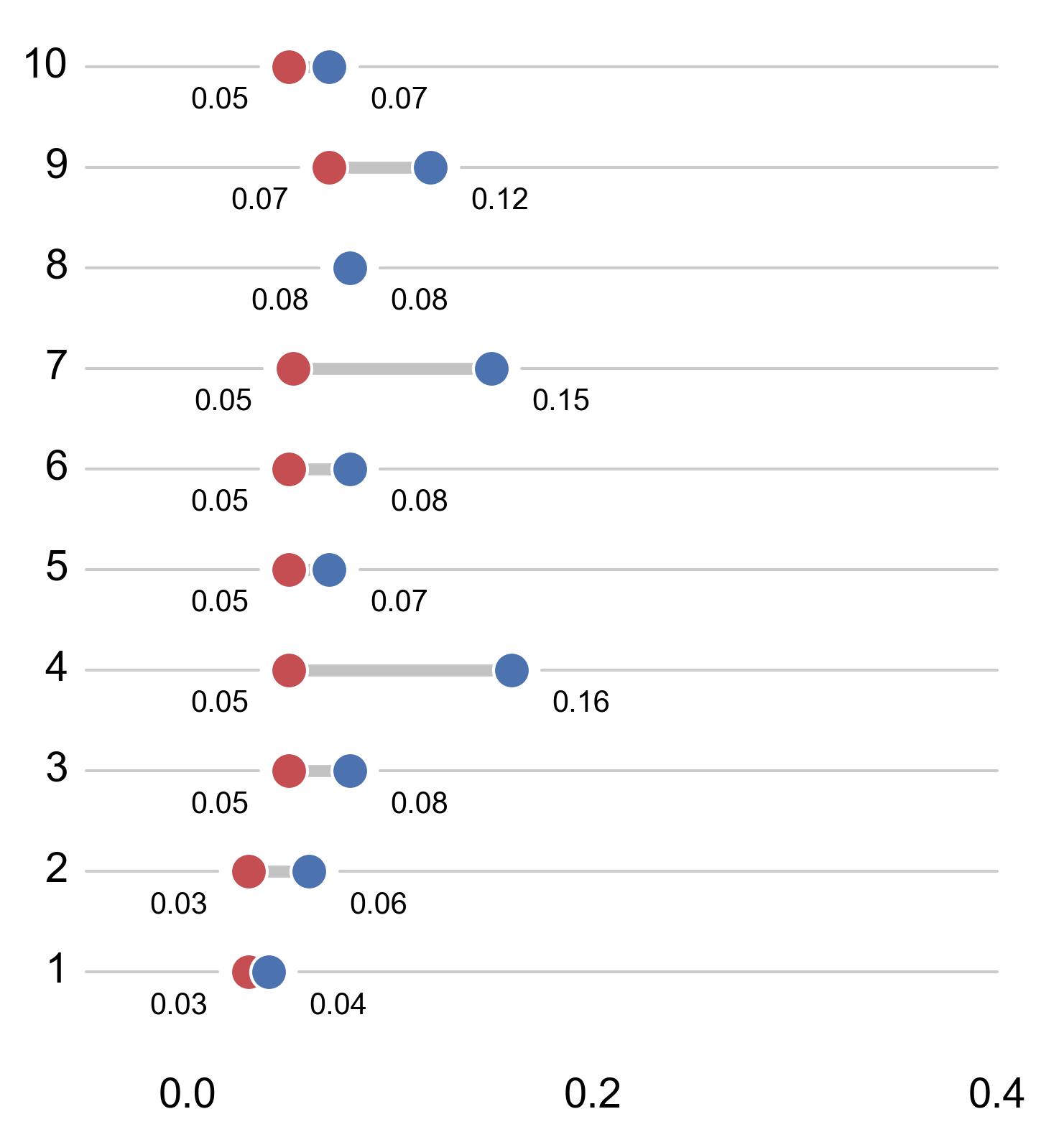}
        \caption{r/War}
        \label{fig:War}
    \end{subfigure}
    \hfill
    \begin{subfigure}[b]{0.45\textwidth}
        \centering
        \includegraphics[width=\textwidth]{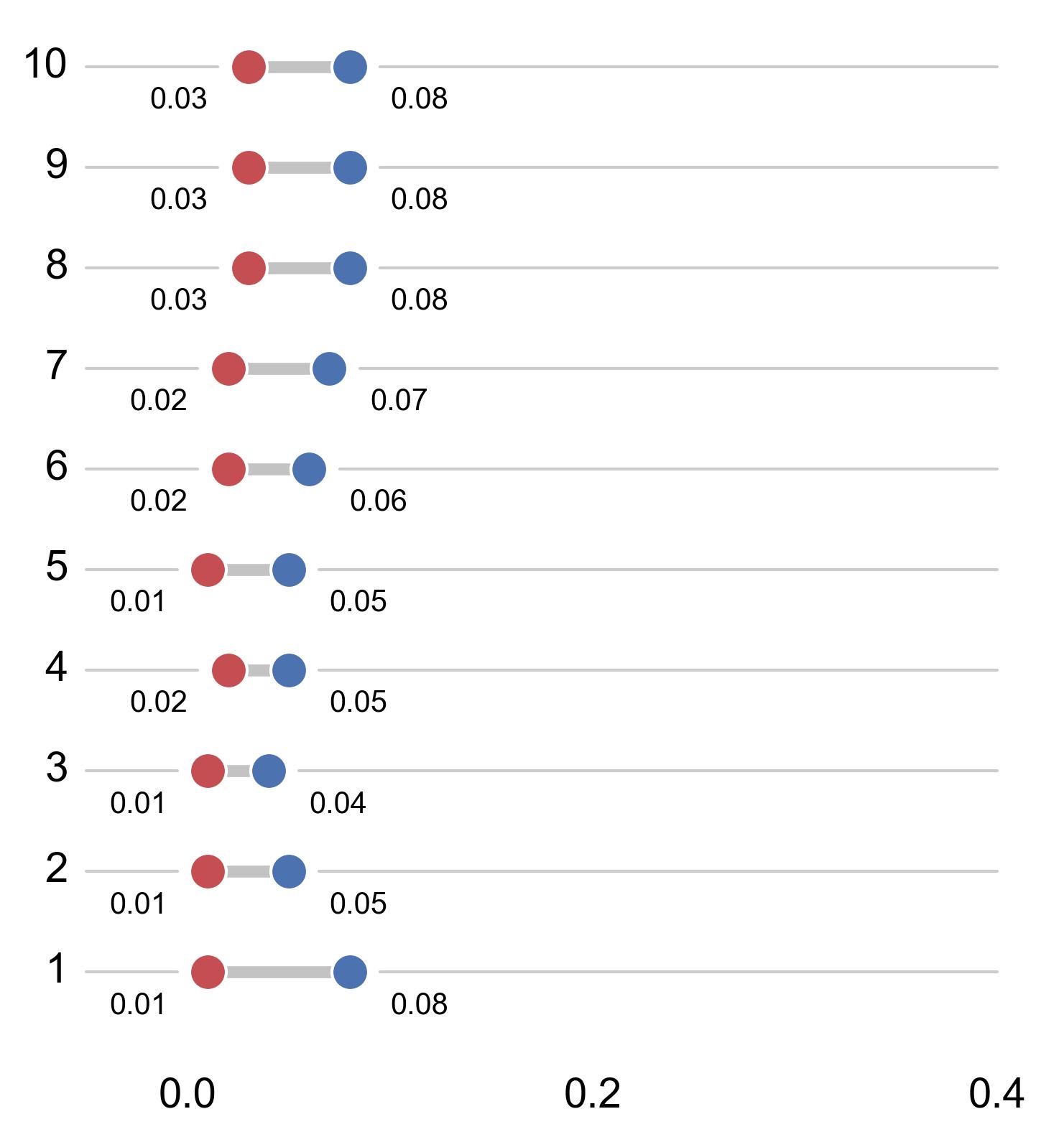}
        \caption{r/Pinoy}
        \label{fig:Pinoy}
    \end{subfigure}
    \vspace{0.5cm}
    
    \begin{subfigure}[b]{0.45\textwidth}
        \centering
        \includegraphics[width=\textwidth]{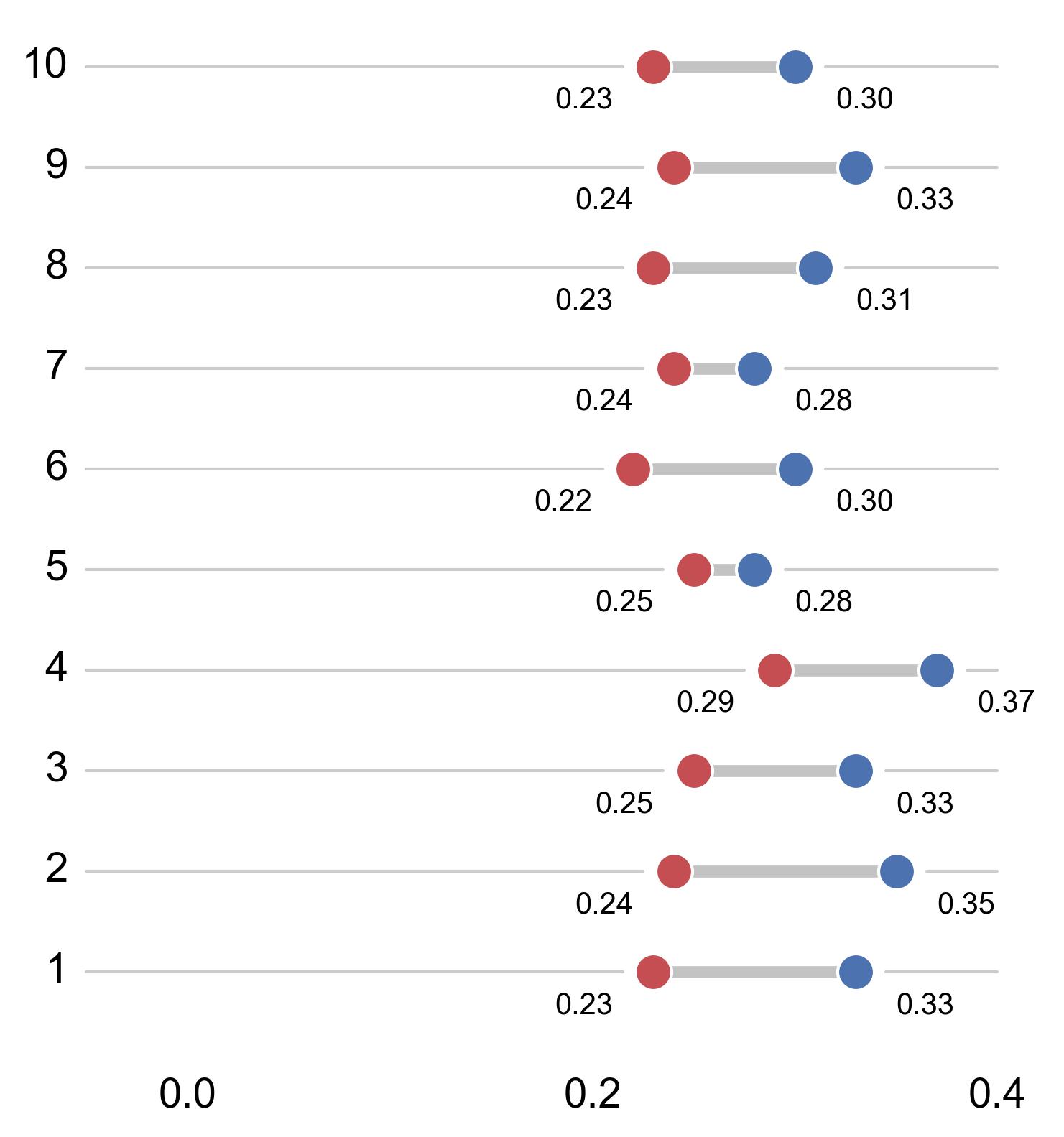}
        \caption{r/Ukraine}
        \label{fig:Ukraine}
    \end{subfigure}
    \hfill
    \begin{subfigure}[b]{0.45\textwidth}
        \centering
        \includegraphics[width=\textwidth]{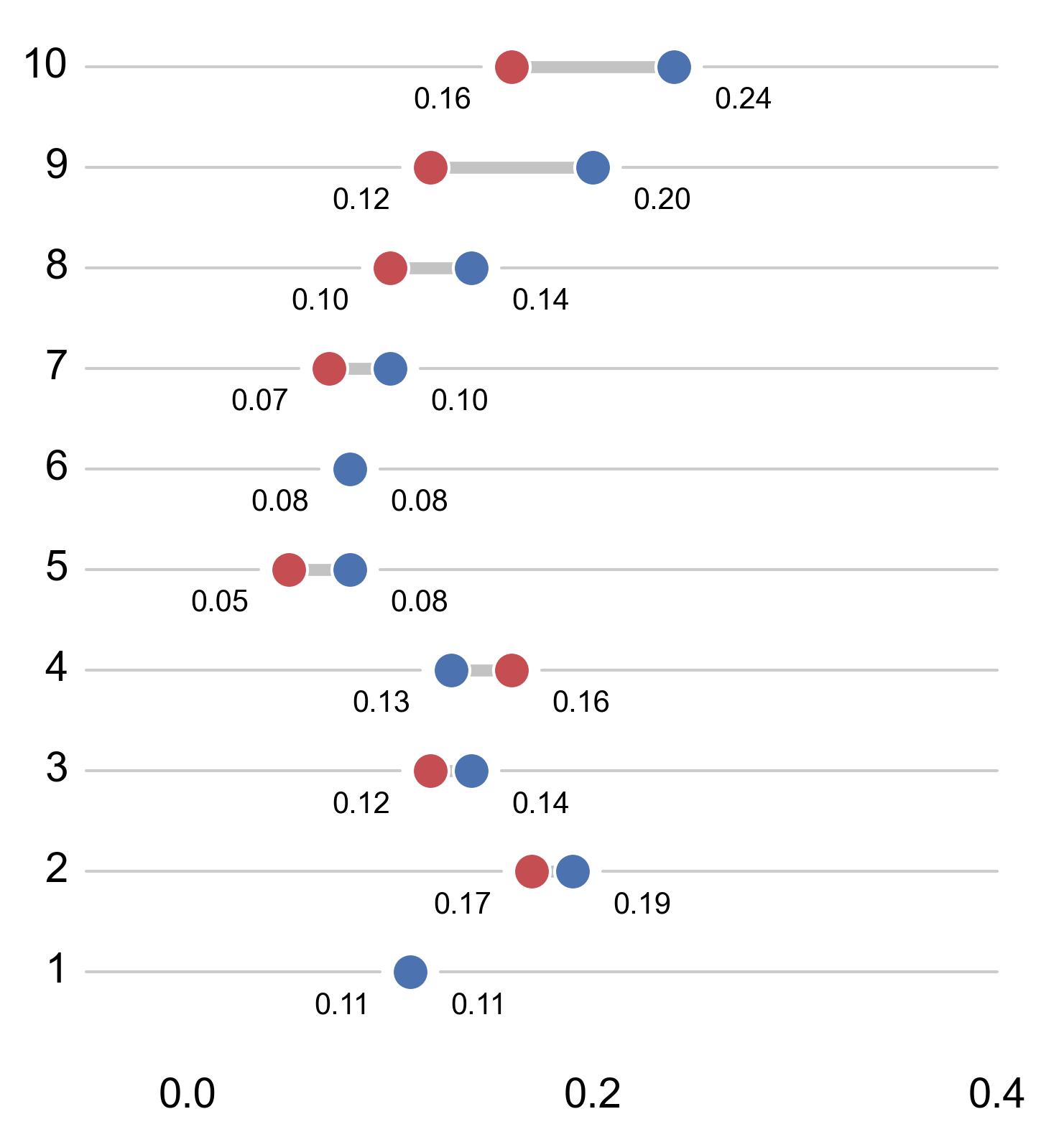}
        \caption{r/Jokes}
        \label{fig:Jokes}
    \end{subfigure}
    
    \caption{Difference in the speed of growth between \blue{\textit{star}} (\blue{blue}) and \blue{\textit{periphery}} (red) subgraphs, averaged over 10 different thread duration (bins) for every subreddit. The speed is computed for every minute. The x-axis represents the number of nodes that join the conversation every minute, while the y-axis represent the length group of threads (10 bins).}
    \label{fig:speeds}
\end{figure}

In order to compare the speed of conversations \blue{with} similar duration \blue{over} time\blue{,} we compute the distribution of the thread length\blue{s} for each subreddit. Then remov\blue{ing} outliers (i.e., extremely long conversations), we group threads by length in time (\blue{dividing them into} 10 bins) and compute the speed for each group of conversations. We calculate the speed of both the two growing subgraphs as follows:

\begin{equation}
    S(g_\blue{m}) = \frac{|e_{\blue{m}}|-|e_{\blue{m}-1}|}{\Delta \blue{m}}
\end{equation}
where $|e_\blue{m}|$ is the total number of edges of the subgraph $g$ at \blue{minute $m$}. We compute the speed for three different time intervals (1 minute, 10 minutes and 1 hour) to observe the growth at different granularities. Speeds that could not be computed because of missing data have been set to 0. Figure \ref{fig:speeds} shows that the difference between the speeds of growth of the two subgraphs is larger in AITA than in other subreddits. The horizontal bars in the plots represent the difference \blue{in} speed as \blue{the} number of nodes that join the conversation every minute. Observe that such \blue{difference} is higher in the AITA community, where the speed of the star \blue{is around 2 and 3 times} the speed of the periphery. \blue{The results for the 10-minute and 1-hour intervals are not plotted for simplicity, as they yield similar results.} We discuss the implication\blue{s} of this result in \blue{Section \ref{sec:conclusions}}. 

\begin{table}[h!]
  \centering
  \small
  \begin{tabular}{ p{2.8cm}  >{\centering}p{2cm} >{\centering}p{1.8cm} p{1.8cm} p{1.6cm} }
  \toprule
  
  \textbf{Subreddit} & \textbf{Threads} & \textbf{\makecell{Avg.\\Comments}} & \textbf{\makecell{\# Data\\collection}} & \textbf{Ref}\\ 
  \cdashline{1-5}
  
  \toprule

r/Geopolitics & 895 & 50 & 2/22 - 5/22& \cite{zhu2022reddit} \\ \cdashline{1-5}
r/War & 4,277 & 23 & 2/22 - 5/22& \cite{zhu2022reddit} \\ \cdashline{1-5}
r/PinoyProgrammer & 3,068 & 15 & 8/23 - 2/24 & \cite{bwandowando_pinoy} \\ \cdashline{1-5}
r/Ukraine & 34,995 & 32 & 12/22 - 1/24 & \cite{bwandowando_ukraine, pohl2023invasion, dvzubur2022semantic} \\ \cdashline{1-5}
r/Jokes & 21,798 & 33 & 5/22 - 12/23 & \cite{bwandowando_jokes} \\

  \bottomrule
  \end{tabular}
  \\[\baselineskip]
  \caption{Data collection of threads from five subreddits.}
  \label{tab:new_datasets_stats}
\end{table}

\subsection{Disagreement and reciprocity}
\label{subsec:reciprocity}

To understand if disagreement in the judgment process \blue{is} what \blue{drives} discussions in AITA conversations, we verify the existence of a monotonic relationship between thread entropy \blue{(computed in Section \ref{subsec:operationalization})} and other features of the threads, \blue{such as: the ASPL and GCC (computed in Section \ref{subsec:results_aita}), the percentage of users participating only once, the length of the thread (in number of comments), the percentage of users that participate without voting, the average length of comments (in number of words), the score of the comments (see Section \ref{subsec:data}), the thread duration over time, the frequency of the comments (number of edges per minute), the average sentiment of the thread, and the percentage of users expressing an ``unsure'' comment (see Section \ref{subsec:temporal_network}). Among these features, we also include a measure of reciprocal interactions.}

\begin{figure}[b!]
    \centering    \includegraphics[width=0.7\textwidth]{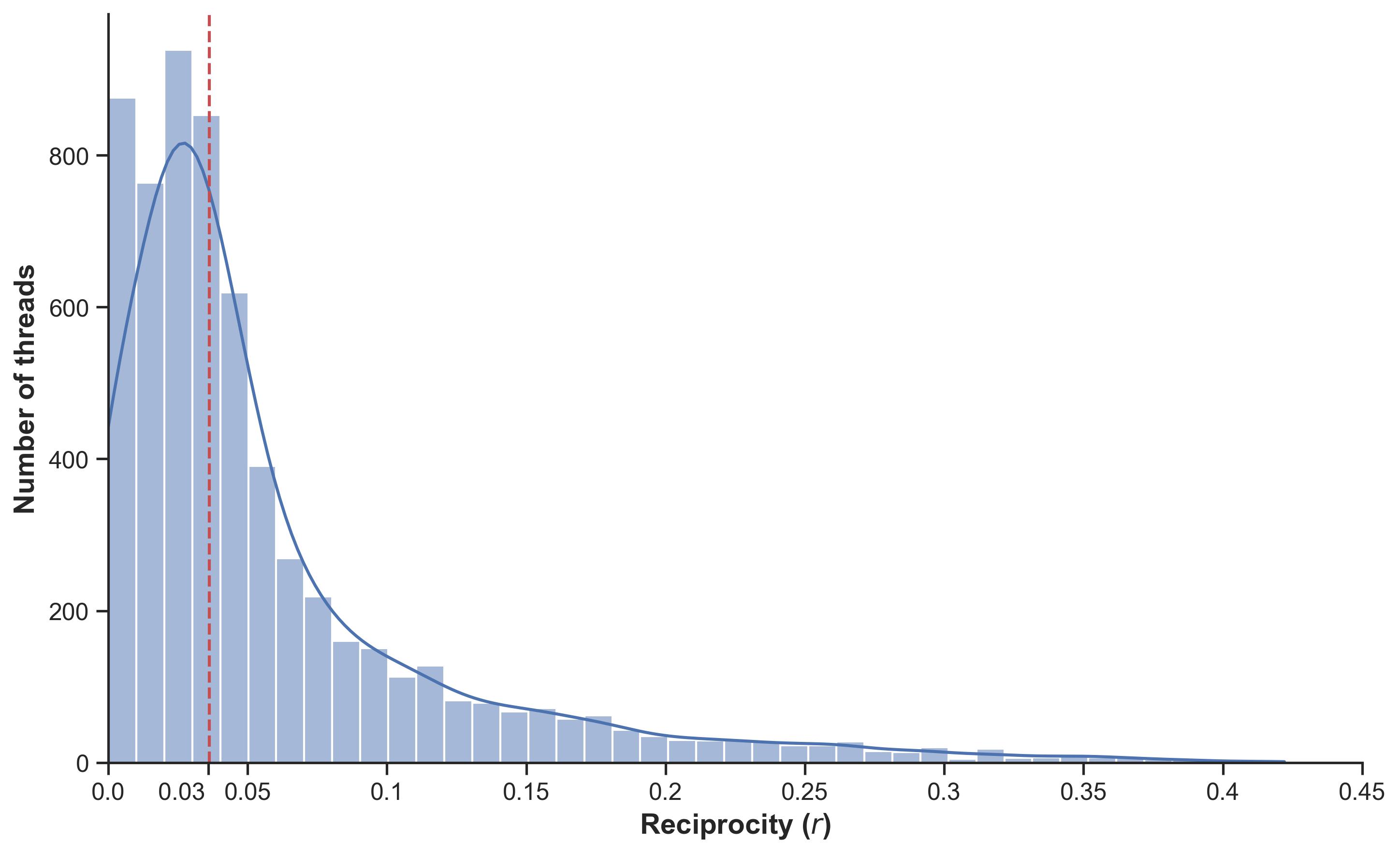}
    \caption{\blue{Distribution of reciprocity ($r$) in AITA networks.}}
    \label{fig:reciprocity_distrib}
\end{figure}

Reciprocity is an important behavioral feature of discussion dynamics that fosters mutual participation in conversations between users~\cite{aragon_thread_2017}\blue{. It} is traditionally deﬁned as
follows~\cite{aragon_thread_2017}: 

\begin{equation}
\label{eq:reciprocity}
    r = \frac{E^{\leftrightarrow}}{E}
\end{equation}

where $E^{\leftrightarrow}$ corresponds to the number of bidirectional edges and $E$ corresponds is the total number of edges. \blue{This metric ranges from 0 to 1, where a value of 0 indicates the absence of reciprocal edges in the network, and a value of 1 indicates that all edges are reciprocated.} We are interested in measuring the amount of reciprocity in AITA threads to assess \blue{its} role in the judgment process, especially \blue{in relation} to disagreement. To characterize reciprocity, we exploit the directed network of replies between users in each thread. In such networks, a directed edge between user $u$ and $v$ exists if user $u$ replied to user $v$ in the discussion. By using th\blue{e} metric \blue{in Equation \ref{eq:reciprocity}}, we compute the reciprocity for every static network. \blue{Figure \ref{fig:reciprocity_distrib} shows the distribution of reciprocity in our dataset of networks. Such distribution is right-skewed, with very small reciprocity for the majority of the threads (0.03 on average), suggesting that very few comments in the AITA discussions are reciprocated. Moreover, while the theoretical upper limit of reciprocity is 1, the maximum value observed for this metric in our data is $\sim0.43$, revealing that there are no threads with high levels of mutual exchange. We interpret such result in connection with other findings in Section \ref{sec:conclusions}.}

\blue{To corroborate the existence of a relationship between disagreement and all the above-mentioned features,} we compute the Spearman rank correlation, with results summarized in Table \ref{tab:correlation} \blue{and discussed in the following Section~\ref{sec:conclusions}}. \blue{We observe that} when thread entropy is high \blue{(}i.e., there is more disagreement in the judgment expressed\blue{),} users \blue{tend to} write more than one comment, often \blue{engaging} in reciprocal discussion\blue{s} with \blue{others}. They also write more comments\blue{,} prefer not to vote,\blue{ and} if they do, they include more than one label in the comment, indicating their uncertainty in picking a side. Not\blue{ably}, when randomizing the networks by edge rewiring, the relation\blue{ship} between disagreement and reciprocal discussions tends to disappear.

\begin{table}[h!]
  \centering
  \small
  \begin{tabular}{ l c }
  \toprule
  
  \textbf{Feature} & \textbf{Coefficient}\\
  \cdashline{1-2}
  
  \toprule

         ASPL & \textbf{0.3} *** \\ \cdashline{1-2}
         GCC & \textbf{0.28} *** \\ \cdashline{1-2}
         only one comment (\%) & \textbf{- 0.38} *** \\ \cdashline{1-2}
         reciprocity & \textbf{0.28} *** \\ \cdashline{1-2}
         reciprocity (rand 20\%) & 0.17 *** \\ \cdashline{1-2}
         reciprocity (rand 50\%) & 0.13 *** \\ \cdashline{1-2}
         reciprocity (rand 90\%) & 0.08 *** \\ \cdashline{1-2}
         comments per thread & \textbf{0.27} *** \\ \cdashline{1-2}
         non-voters (\%) & \textbf{0.4} *** \\ \cdashline{1-2}
         comment length (avg) & 0.005 \\ \cdashline{1-2}
         comment score (avg) & 0.06 *** \\ \cdashline{1-2}
         thread duration & 0.17 *** \\ \cdashline{1-2}
         comment frequency & 0.14 *** \\ \cdashline{1-2}
         post sentiment & 0.05 *** \\ \cdashline{1-2}
         unsure voters & \textbf{0.48} ***\\

  \bottomrule
  \end{tabular}
  \\[\baselineskip]
  \caption{Spearman rank correlation between disagreement (thread entropy) and thread features. Asterisks mark significance levels (*** p-value$<$.001)}
  \label{tab:correlation}
\end{table}

\section{Discussion and conclusion}
\label{sec:conclusions}

In this paper\blue{,} we analyzed Reddit threads by modeling them as networks of user interactions and by computing the evolution of their structural properties \blue{over} time. We show that \blue{these} networks differ from real social networks, despite falling in the same category, as they exhibit a negligible GCC and an increasing ASPL. We also \blue{demonstrated} that networks of the AITA community grow differently with respect to networks from other subreddits\blue{, as} the difference \blue{in} speed between the two subgraphs is larger than in other subreddits. In this section\blue{,} we discuss such results \blue{in the context} o\blue{f} Social Judgement Theory, \blue{particularly} regard\blue{ing} disagreement in the judgment process. 

We interpret the results presented in \blue{Section~\ref{subsec:results_aita}} \blue{by} referring to the structure of the platform, \blue{which} allow\blue{s} threaded-structured conversations \blue{and} shapes the user interaction differently \blue{compared} to other real social networks. Indeed, Reddit is not a relationship-based social network, meaning that most of the user interactions are content-driven and not user-driven~\cite{makow_2017}. This means that users \blue{on} Reddit do not join to comment \blue{on} a specific person but \blue{on} a specific content (post or comment). \blue{This difference} in how the platform is built shapes user interactions differently, generating \blue{a} different behavior \blue{in} the networks \blue{as they} evolv\blue{e} over time. 

Furthermore, the unexpected behavior of GCC and ASPL reveals that participants mostly interact with only one other user, and often by a single message\blue{.} To further \blue{inspect such user behavior}, we derive\blue{d} a measure of reciprocal interaction \blue{and its relation to the disagreement in the judgment process of AITA threads}. We \blue{have} show\blue{n} \blue{in Section~\ref{subsec:reciprocity}} that disagreement plays an important role in online discussions where people are expected to express a judgment\blue{. I}t is significantly related to the generation of more discussions and more reciprocal interactions and, at the same time, to more uncertainty in judgment expression. This could reveal that, despite the anonymity of users on Reddit, users might not feel free \blue{to} explicitly express their opinion\blue{s} in discussions with high disagreement. \blue{This is coherent with SJT: indeed, if it is true that people are more prone to express opinions in anonymous environments and settings \cite{Adamic_2021, spears_social_2021}, it is also true that in situations of high disagreement they perceive less support for their viewpoint from the social environment, \blue{making them} less likely to express their judgments~\cite{glynn_97, CHUN2017120}.
Furthermore, in relation to the expression of social judgment in online discussions, in this work we have also shown that comments containing a judgment have a higher response time than comments that do not include it (Section \ref{subsec:results_substructures}). This finding aligns with moral judgment theories stating that responses to moral dilemmas require cognitive control, which is an emotional process that takes time~\cite{suter_2011}. The more time needed for voting comments could also be due to the AITA community guidelines that encourage users to include a justification for the expressed vote in the text of their comments.
In summary, the obtained results contribute to the advancement of unexplored aspects of the SJT, especially related to online communication.} 

We conclude that the temporal analysis of the structural properties of \blue{these} networks reveals the \blue{following} behavioral patterns of users discussing \blue{on Reddit}. \blue{P}articipants mostly interact with only one other user, often by a single message. The lack of clusters, together with the very small reciprocity, suggests that most of the new users participating in the conversation do not engage with more than one person. They join the thread to \blue{respond} to a single user, rarely with more than one message exchange. 

\blue{In this work we also} demonstrate that the speed of the star in the AITA conversations grows faster than the periphery \blue{(Section \ref{subsec:results_reddit})}. We interpret this as a consequence of community guidelines enforcing the behavior of participants, since it is a direct consequence of the community rules. As explained in Section \ref{subsubsec:judgment}, these rules indicate that votes expressed in comments that are not first-level will not be considered for the final judgment verdict, hence encouraging people to participate in the thread by answering to the post author. The periphery is, as a consequence, a spontaneous behavior of the users who discuss instead of voting (only 30\% of the voters are voting in the periphery, as shown in Figure \ref{fig:user_distribution}).

\section*{Availability of data and materials}
The datasets analyzed during the current study are available from the corresponding author on reasonable request.

\section*{Competing interests}
The authors declare that they have no competing interests.

\section*{Funding}
This work has been partly funded by eSSENCE, an e-Science collaboration funded as a strategic research area of Sweden. The funders had no role in study design, data collection and analysis, decision to publish, or preparation of the manuscript.

\section*{Author's contributions}
Both authors contributed to the design and implementation of the research. The data acquisition, data curation, formal analysis and validation has been carried by D.G. Both authors contributed to the analysis of the results and to the writing of the manuscript. Both authors read and approved the final manuscript.

\section*{Acknowledgements}
We would like to thank Prof. Matteo Magnani for his comments and suggestions.


\printbibliography[heading=bibintoc]

%
%
%



\end{document}